\documentclass[useAMS,usenatbib]{mn2e}
\pdfoutput=1
\usepackage{aas_macros}
\usepackage{epsfig}
\usepackage{amsmath}
\usepackage{amssymb}
\usepackage{caption}
\usepackage{leftidx}
\usepackage{natbib}
\usepackage[rgb]{xcolor}
\definecolor{MyGreen}{rgb}{0.0,0.6,0.3}
\definecolor{MyPurple}{rgb}{0.6,0,0.3}
\usepackage{fancyref}
\usepackage{hyperref}
\hypersetup{colorlinks=true,citecolor=MyGreen,linkcolor=MyPurple,urlcolor=blue}

\def\beq{\begin{equation}}
\def\eeq{\end{equation}}
\def\ba{\begin{eqnarray}}
\def\ea{\end{eqnarray}}
\def\bal{\begin{align}}
\def\eal{\end{align}}
\def\bxi{{\mbox{\boldmath $\xi$}}}

\begin{document}

\title[Pre-Supernova Outbursts] {Pre-Supernova Outbursts via Wave Heating in Massive Stars II: Hydrogen-poor Stars}

\author[Fuller \& Ro]{
Jim Fuller$^{1,2}$\thanks{Email: jfuller@caltech.edu},
Stephen Ro$^{3}$\\
\\$^1$TAPIR, Walter Burke Institute for Theoretical Physics, Mailcode 350-17, Caltech, Pasadena, CA 91125, USA
\\$^2$Kavli Institute for Theoretical Physics, Kohn Hall, University of California, Santa Barbara, CA 93106, USA
\\$^3$Astronomy Department, University of California at Berkeley, Berkeley, CA 94720, USA
}

\label{firstpage}
\maketitle

\begin{abstract}

Pre-supernova (SN) outbursts from massive stars may be driven by hydrodynamical wave energy emerging from the core of the progenitor star during late nuclear burning phases. Here, we examine the effects of wave heating in stars containing little or no hydrogen, i.e., progenitors of type IIb/Ib SNe. Because there is no massive hydrogen envelope, wave energy is thermalized near the stellar surface where the overlying atmospheric mass is small but the optical depth is large. Wave energy can thus unbind this material, driving an optically thick, super-Eddington wind. Using 1D hydodynamic MESA simulations of $\sim \! 5 \, M_\odot$ He stars, we find that wave heating can drive pre-SN outbursts composed of a dense wind whose mass loss rate can exceed $\sim \! 0.1 \, M_\odot/{\rm yr}$. The wind terminal velocities are a few $100 \, {\rm km}/{\rm s}$, and outburst luminosities can reach $\sim \! 10^6 \, L_\odot$. Wave-driven outbursts may be linked with observed or inferred pre-SN outbursts of type Ibn/transitional/transformational SNe, and pre-SN wave-driven mass loss is a good candidate to produce these types of SNe. However, we also show that non-linear wave breaking in the core of the star may prevent such outbursts in stars with thick convective helium-burning shells. Hence, only a limited subset of SN progenitors are likely to experience wave-driven pre-SN outbursts.

\end{abstract}


\section{Introduction}

The heterogeneity of core-collapse supernovae (SNe) reveals the great diversity of massive star SN progenitors approaching death. Much of this diversity has been theoretically anticipated: different progenitor masses, metallicities, and rotation rates produce substantial diversity. More importantly, binary interactions can drastically alter the stellar structure at death, as the majority of hydrogen-poor SNe (types IIb, Ib, Ic, etc.) are widely considered to be the result of hydrogen envelope stripping by a companion star.

Despite the theoretically expected diversity of core-collapse SNe, many aspects of this diversity are unexpected even in binary stellar evolution models. Of particular interest are the growing number of hydrogen-poor SNe that exhibit evidence of enhanced pre-SN mass loss or outbursts, such as type Ibn SNe \citep{pastorello:08,pastorello:16,hosseinzadeh:17} which show interaction with He-rich material ejected soon before core-collapse. Specific examples of Ibn SNe include SN 2006jc (which had a pre-SN outburst, \citealt{pastorello:07}), SN 2015G \citep{shivvers:17}, and SN 2015U, \citep{shivvers:16}. PTF11qcj was as Type Ic SN \citep{corsi:14} which may have also exhibited a pre-SN outburst and enhanced mass loss.

Some SNe also show evidence for small amounts of hydrogen ($\lesssim \! 1 \, M_\odot$) ejected in the years or decades prior to the SNe. Examples include transitional IIn/Ibn SNe such as SN 2005la \citep{pastorello:08b} and SN 2011hw \citep{smith:12}. The transformational SN 2014C appeared as a relatively normal type Ib SNe, until it transitioned into a type IIn SNe when its ejecta ran into  H-rich CSM ejected $\sim$decades before explosion \citep{milisavljevic:15,margutti:16}. Early spectra of type IIb SN 2013cu reveal emission lines from a flash-ionized wind \citep{gal-yam:14} with inferred mass loss rates over $10^{-3} \, M_\odot/{\rm yr}$ \citep{groh:14}, and total CSM mass of at least $0.3 \, M_\odot$ \citep{grafener:16}.

In a previous paper \citep{fuller:17}, we examined the possibility of pre-SN outbursts in hydrogen-rich red supergiants (RSGs). In that work, we investigated the role of energy transported by convectively excited hydrodynamic waves during late stage nuclear burning in the stellar core, an idea originally proposed by \cite{quataert:12}, with follow-up work in \cite{shiode:14}. We found wave energy transport could transport $\sim \! 10^7 \, L_\odot$ of power from the core to the envelope during core Ne/O burning, enough to power a pre-SN outburst and enhanced mass loss. In RSGs with thick hydrogen envelopes, waves thermalize their energy just outside the helium core at the base of the hydrogen envelope. The hydrogen envelope is too massive to be totally ejected, and it prevents most wave heat from diffusing outward, such that the observed pre-SN outburst is fairly mild.

Here, we examine the effects of wave energy transport in stars lacking thick hydrogen envelopes, which greatly alters the effects of wave heating.  In hydrogen-poor progenitor stars, we find that waves thermalize much closer to the stellar surface, such that the binding energy of the overlying atmosphere is much smaller than the wave energy budget. The wave heat can thus power a dense, super-Eddington wind as discussed in \cite{quataert:16,shen:16,owocki:17}. Although the emergent radiative luminosity of this wind is super-Eddington and is larger than in the RSG case, most of the wave energy is used to unbind mass and accelerate it above the escape velocity, with only a small fraction escaping as radiation.

%Our paper is organized as follows. \autoref{imp} examines wave excitation, propagation, dissipation, and implementation of wave heating in the models. In \autoref{effects}, we describe the effects of wave heating on our stellar models. We describe the implications for pre-SN outbursts and subsequent in SNe in \autoref{disc}, and we also discuss the limitations of our models. \autoref{conclusions} presents a summary of our results.

\section{Basic Idea}

The evolution of massive stars is complex, especially near the end of their lives when evolutionary timescales become short. Here we review the end stages of massive star evolution (see \citealt{woosley:02}), and discuss why wave energy transport is likely to be important as proposed by \cite{quataert:12}.

By the onset of core carbon burning, core temperatures and densities are high enough that neutrino cooling (rather than convection or radiative diffusion) carries away most of the energy generated by nuclear reactions. While neutrino cooling is strongly sensitive to temperature (roughly $\epsilon_{\nu,{\rm cool}} \propto T^{10}$), late-stage nuclear burning has even stronger dependence (roughly $\epsilon_{\rm nuc} \! \propto \! T^{40}$). Nuclear burning outweighs neutrino cooling at the center of the star where the temperatures are highest, and the strong temperature gradient generates convection. At the outer edge of the convection zones, neutrino cooling outweighs nuclear burning, eventually carrying away most of the generated energy. Nonetheless, convection persists below and locally carries nearly all the generated nuclear energy. Because neutrino cooling allows the core to cool quickly, nuclear burning proceeds on neutrino cooling timescales, and core convective luminosities may exceed the surface luminosity of the star by several orders of magnitude.

The key insight of \cite{quataert:12} is that convection always excites internal gravity waves (IGW) at the interface with a radiatve zone, as verified by numerous simulations \citep{meakin:06,dessart:06,meakina:07,meakinb:07,rogers:13,alvan:14,alvan:15,lecoanet:15,cristini:17,rogers:17}. Although only a small fraction of the convective flux is converted into IGW, the extreme convective luminosities of late-stage nuclear burning can excite IGW carrying $\sim \!  10^7 \, {\rm L_\odot}$, much more power than the star's surface luminosity. The IGW transport their energy on wave crossing timescales, which are comparable to dynamical timescales. In the absence of dissipative effects, the wave action is conserved, and waves can potentially carry their energy over great distances before depositing it where they damp.

\begin{figure}
\begin{center}
\includegraphics[scale=0.35]{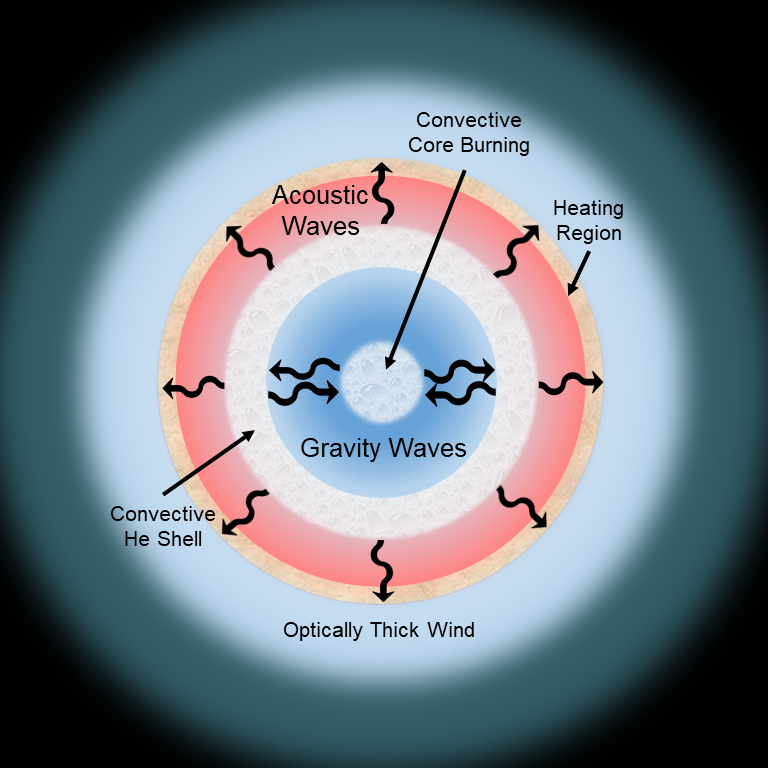}
\end{center} 
\caption{ \label{fig:Cartoon} 
Cartoon (not to scale) of wave heating in a hydrogen-poor star. Gravity waves are excited by vigorous core convection and propagate through the outer core. After tunneling through the evanescent region created by the convective helium-burning shell, they propagate into the envelope as acoustic waves, which damp below the photosphere and heat a thin shell. The intense wave heating drives a super-Eddington wind.}
\end{figure}

The effects of the wave energy transport thus depend on where IGW eventually damp. The IGW redistribute little energy ($E_{\rm IGW,tot} \sim \! 10^{48} \, {\rm erg}$) relative to the helium core binding energy. If they dissipate within the core, their energetic effect on the star's evolution will be negligible (additionally, wave energy thermalized in the core will be lost through neutrino cooling). However, the wave luminosity can exceed both the surface luminosity and the Eddington luminosity, while their total energy can be much larger than the envelope binding energy. Consequently, if the waves can dissipate their energy in the star's envelope, their effect can be dramatic. As we show below, wave energy thermalized in the envelopes of hydrogen-poor stars will drive an optically thick super-Eddington outflow. This outflow could be detected as a pre-SN outburst, and interaction between the SN ejecta and the pre-SN outflow can alter the appearance of the SN, evidenced by the observations discussed above. Figure \ref{fig:Cartoon} illustrates the basic process of wave energy transport.

\section{Implementing Wave Energy Transport in Stellar Models}
\label{imp}

\subsection{Stellar Models}
\label{mods}

Our implementation of wave energy transport closely follows the methods of \cite{fuller:17}, using the stellar evolutin code MESA \citep{paxton:11,paxton:13,paxton:15}. The first step is to create hydrogen-poor stellar models. We use the same $15 \, M_\odot$ stellar model of \cite{fuller:17}, however, once the star completes core helium burning, we simulate binary mass loss by stripping off nearly the entire hydrogen envelope (details in \autoref{models}). This leaves a nearly hydrogen-free star with a helium core mass of $M_{\rm He} \simeq 5.3 \, M_\odot$. We remove the hydrogen envelope after core He burning (mimicking case C mass transfer) so that the helium core mass is nearly the same as the model of \cite{fuller:17}. This results in nearly identical subsequent core evolution and wave energetics, for easier comparison with the hydrogen-rich models of \cite{fuller:17}.

In this paper, we examine two stellar models. The first is completely hydrogen-free, and likely represents a fairly typical type Ib SN progenitor formed via binary interaction. At the onset of carbon burning, it has a radius of $R \! \sim 4 \! \, R_\odot$ and surface temperature $T_{\rm eff} \sim 5 \! \times \! 10^4 \, {\rm K}$. The second model contains a light but inflated hydrogen envelope of $M_{H} \sim 3 \! \times \! 10^{-2} \, M_\odot$. At the start of carbon burning, it has a radius of $R \sim 450 \, R_\odot$ and surface temperature of $T_{\rm eff} \sim 5000 \, {\rm K}$. It likely represents a fairly typical (perhaps more extended and cooler than average) type IIb SN progenitor.

\begin{table}
\caption{Model Timescales} \label{table} 
\begin{tabular}{|c|c|c|c|c|}
\hline 
Burning Phase & $t_{\rm col}$ & $t_{\rm dyn}$ & $t_{\rm therm}$ & $t_{\rm con}$ \\ 
\hline 
C-burning & 300 yr & 5 s & $10^3 \, {\rm yr}$ & $6 \! \times \! 10^4 \, {\rm s}$ \\ 
\hline 
O-burning & 0.2 yr & 1 s & 0.1 yr & 900 s \\ 
\hline 
Si-burning & 2 days & 0.5 s & 3 days & 200s \\ 
\hline 
\end{tabular} 
\\ Time until core-collapse $t_{\rm col}$, local dynamical time $t_{\rm dyn} = H/c_s$, local thermal time $t_{\rm therm} = 4 \pi \rho H c_s^2/L$, and convective turnover time $t_{\rm con} = 2 \pi/\omega_{\rm con}$, evaluated in the middle of core convective zones during burning phases as labeled.
\end{table}

Although our stellar models are not perfect, several lines of evidence suggest they capture the correct basic stellar structure. First, stellar evolution models of moderate-mass $M \lesssim 20 \, M_\odot$ stars are generally successful at predicting basic stellar evolution and stellar structure (as verified by asteroseismic and countless other observations). Second, while different massive stellar evolution codes (e.g., MESA, Kepler, Geneva) differ somewhat in their treatment of mixing and mass loss, they all agree on the basic core structure and evolutionary sequence of moderate-mass SN progenitors \citep{woosley:02,meakin:11}. It is this core structure, such as the luminosity and extent of convective C/O/Ne burning zones, that is important here. Finally, multi-dimensional simulations of late-stage burning in massive stars (\citealt{meakin:06,meakina:07,couch:15,lecoanet:16,muller:16,jones:17,cristini:17}) do not exhibit core structure or evolution drastically different from one-dimensional models.
%Though these simulations do not run long enough to capture effects on long timescales, they do not indicate any dynamical timescale instabilities in the final stages of evolution.
Even in the final years of a massive star's life, evolutionary timescales are much longer than the core dynamical timescale (hydrostatic equilibrium is maintained), convective speeds remain highly subsonic, and one-dimensional models are a reasonable approximation of the basic structure.  Table \ref{table} compares remaining lifetimes, dynamical times, thermal times, and convective turnover times for our models during a few different burning stages.

Beginning at core carbon burning, we include wave heating effects as described below. Additionally, we utilize MESA's one dimensional implicit hydrodynamics capability to capture the wave-driven wind (see \autoref{models}). Our outer boundary is placed at an optical depth of $10^{-2}$, at which point outflowing material is removed from the grid, and our mass loss rates are defined as the mass flux through this outer boundary.

\subsection{Wave Generation}
\label{gen}

We add wave energy transport to our models in almost the same way as \cite{fuller:17}. At each time step, we compute the wave energy flux generated by the innermost convective zone of the star via
\beq
\label{Lwave}
L_{\rm wave} \sim \mathcal{M}_{\rm con} L_{\rm con} \, ,
\eeq
with $L_{\rm con}$ is the convective luminosity,  and $\mathcal{M}_{\rm con} = v_{\rm con}/c_s$ is the mixing length theory (MLT) convective Mach number, and $c_s$ is sound speed. The MLT convective velocity and turnover frequencies are calculated via
\beq
\label{vconv}
v_{\rm con}=\big[L_{\rm con}/(4 \pi \rho r^2)\big]^{1/3} \, ,
\eeq
and
\beq
\label{omegaconv}
\omega_{\rm con} = 2 \pi \frac{v_{\rm con}}{2 \alpha_{\rm MLT} H} \, ,
\eeq
with $\alpha_{\rm MLT} H$ the mixing length (see discussion on convection in \citealt{meakina:07,alvan:14,couch:15,lecoanet:16,jones:17,cristini:17}). As in \cite{fuller:17}, we assume dipolar ($l=1$) waves with angular frequency 
\beq
\label{omegawave}
\omega_{\rm wave} = \omega_{\rm con,max} \, ,
\eeq
where $\omega_{\rm con,max}$ is the maximum value obtained within a convective zone (for core convection, we find $\omega_{\rm con}$ does not vary much within a convective zone). Our implementation of waves is a drastic simplification (see simulations of, e.g., \citealt{rogers:13,rogers:15,alvan:14,alvan:15}), as discussed in \cite{fuller:17}, but represents a first step in implementing wave energy transport into a stellar evolution code.

\subsection{Wave Propagation}
\label{prop}

If waves remain linear, their propagation through a star can be very accurately calculated, as demonstrated by the remarkable successes of asteroseismology and helioseismology. Linear wave propagation can be understood from the wave dispersion relation
\beq
k_r^2 = \frac{ \big(N^2 - \omega_{\rm wave}^2 \big) \big(L_\ell^2 - \omega_{\rm wave}^2 \big) }{\omega_{\rm wave}^2 c_s^2} \, ,
\eeq
where $k_r$ is the radial wavenumber, $N^2$ is the Brunt-V\"{a}is\"{a}l\"{a} frequency squared, and $L_\ell^2 = \ell(\ell+1) c_s^2/r^2$ is the Lamb frequency squared. In the limit that $\omega_{\rm wave} \ll N, L_\ell$, the dispersion relation reduces to that of IGW, 
\beq
k_{r,{\rm IGW}}^2 \simeq \frac{ \ell (\ell+1) N^2 }{\omega_{\rm wave}^2 r^2}  ,
\eeq
whose group velocity is $v_{\rm IGW} \simeq  \omega_{\rm wave}^2 r/\sqrt{l(l+1)N^2}$. In the limit that $\omega_{\rm wave} \gg N, L_\ell$, the dispersion relation reduces to that of acoustic waves, 
\beq
k_{r,{\rm ac}}^2 \simeq \frac{\omega_{\rm wave}^2}{c_s^2} \, ,
\eeq
whose group velocity is $v_{\rm ac} \simeq c_s$. In regions of the star where one of these two criteria are satisfied, waves propagate freely, conserving their energy apart from weak damping effects.

\begin{figure}
\begin{center}
\includegraphics[scale=0.35]{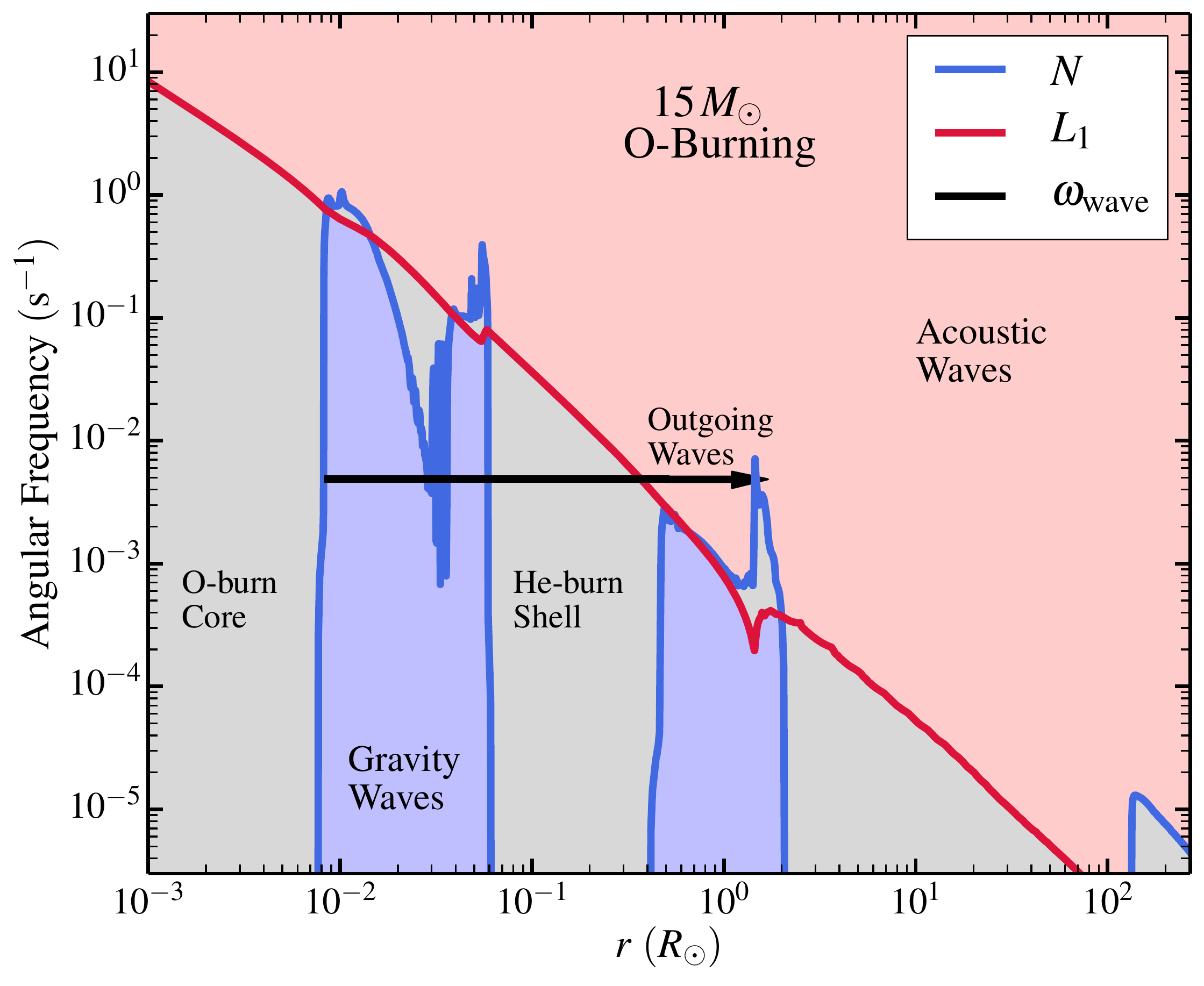}
\end{center} 
\caption{ \label{fig:15MsunProp_stripped} 
Propagation diagram for our hydrogen-free stellar model during core oxygen burning, showing the Brunt-V\"{a}is\"{a}l\"{a} frequency $N$ and the $\ell=1$ Lamb frequency $L_1$. Core convection excited waves at angular frequencies $\omega_{\rm con} \sim 5 \times 10^{-3} \, {\rm s}^{-1}$ that propagate as gravity waves between the oxygen-burning core and the helium-burning shell. They tunnel into acoustic waves above the helium-burning shell and damp in the helium envelope at $\sim \! 2 \, R_\odot$. Above this region, the envelope consists of an optically thick outflowing wind.}
\end{figure} 

In the limit that $N > \omega_{\rm wave} > L_\ell$ or $L_\ell > \omega_{\rm wave} > N$, the radial wavenumber is imaginary. The waves become evanescent in these regions of the star, with the wave amplitude exponentially decreasing over a skin depth $l = |k_r|^{-1} \simeq r/\sqrt{\ell(\ell+1)}$ in the latter case. This does not mean the waves are damped, but rather that they are reflected from surfaces where $\omega_{\rm wave}=N$ or $\omega_{\rm wave}=L_\ell$. However, a fraction of wave energy will tunnel through this evanescent barrier, with the fraction of transmitted wave energy given by 
\beq
\label{trans}
T^2 = \exp \bigg(- 2 \int^{r_1}_{r_0} \frac{dr}{l} \bigg) \approx \bigg(\frac{r_0}{r_1}\bigg)^{2 \sqrt{\ell(\ell+1)}} \, .
\eeq
Here, $r_0$ and $r_1$ are the boundaries of the evanescent region where $\omega_{\rm wave} =N$ or $L_\ell$. This transmission of wave energy through evanescent regions is well-understood  in low-mass red giant stars (see discussion in \citealt{fuller:15,takata:16b,mosser:17}) in which mixed acoustic-gravity modes tunnel through an evanescent layer separating the radiative core from the convective envelope. There is no reason to suspect the process will operate differently in massive stars, unless the waves are very non-linear or the core's structure is very different than predicted by models.

\autoref{fig:15MsunProp_stripped} shows a wave propagation diagram for our hydrogen-free model during core oxygen burning. Waves excited at convective turnover frequencies $\omega_{\rm wave} \! \sim \! 5 \times 10^{-3} \, {\rm rad}/{\rm s}$ propagate as IGW within the He/C/O core, and acoustic waves in the stellar envelope. To penetrate into the envelope, the waves must tunnel through the intervening evanescent region created by the convective helium-burning shell. The structure is similar to the hydrogen-rich case, which also contains a region near the top of the helium-burning shell where gravity waves transition into acoustic waves. Note that \autoref{fig:15MsunProp_stripped} is shown during oxygen burning, after waves have initiated a strong outflow. The stellar envelope is more compact during prior evolution before the onset of strong wave heating during late burning phases. Once the wave heating rate is significantly super-Eddington, we find the stellar structure generally consists of a hydrostatic core below the wave heating region, surrounded by an optically thick supersonic outflow above the wave heating region.

In our models, we calculate the transmission coefficients of each evanescent region between the core and envelope by performing the integral of equation \ref{trans}. We also calculate the wave damping rate in the core due to neutrino losses and wave-breaking at the center of the stars, and we then calculate the fraction of wave energy that escapes the core to heat the envelope,
\beq
\label{fesc}
f_{\rm esc} = \bigg[1 + \frac{f_{\rm damp}}{T_{\rm min}^2} \bigg]^{-1} \,
\eeq
where $f_{\rm damp}$ is the fraction of wave energy that damps in one wave crossing time of the core (see \citealt{fuller:17}).

\subsection{Wave Dissipation via Weak Shocks}
\label{shocks}

Waves that manage to tunnel into the envelope as acoustic waves will then propagate toward the stellar surface where they become strongly damped. \cite{ro:17} showed that acoustic waves carrying more energy than the background radiation field will generally steepen into shocks before dissipating by radiative diffusion. Here we describe how we account for shock formation and entropy generation in our models. The full implementation is discussed below, but typically we find that waves damp just below the point where their energy flux $L_{\rm waves}$ approaches the maximum possible wave flux (in the linear regime)
\beq
\label{Lmax}
L_{\rm max} = 2 \pi r^2 \rho c_s^3 \, .
\eeq

The first step is to calculate the wave amplitude based on the wave energy flux. In the envelope of the star where the waves are acoustic waves and have not yet damped or shocked, their local amplitude can be calculated from WKB scaling relations
\beq
\label{LWKB}
L_{\rm waves} \simeq 2 \pi r^2 \rho c_s u^2 \simeq {\rm constant} \, ,
\eeq
where $u$ is the radial velocity amplitude of the wave. We then use equation 7 of \cite{ro:17} to calculate the radial coordinate $r_S$ at which the waves form shocks,
\begin{align}
\label{rshock}
& \bigg(\frac{ \partial u}{\partial r} \bigg)_{\! \! i} \int^{r_S}_{r_i} dr \frac{\gamma+1}{2 c_s} \bigg( \frac{L_{\rm max}(r_i)}{L_{\rm max}} \bigg)^{1/2} = 1 \nonumber \\  
& \simeq \frac{\omega \, u(r_i) }{c_s(r_i)} \int^{r_S}_{r_i} dr \frac{\gamma+1}{2 c_s} \bigg( \frac{L_{\rm max}(r_i)}{L_{\rm max}} \bigg)^{1/2} = 1 \, .
\end{align}
Here, we have used the WKB dispersion relation for sound waves, $k_r \approx \omega/c_s$, and $\gamma$ is the adiabatic index. The integral in equation \ref{rshock} is taken from the radius $r_i$ at which acoustic waves are formed, and the shock formation radius $r_S$ is defined by the radial coordinate at which the left hand side of equation \ref{rshock} evaluates to 1. We evaluate $r_i$ by finding the deepest point in the star where $\omega > L_l,N$ such that the waves behave as acoustic waves. In practice, the value of $r_i$ is not important because the integral of equation \ref{rshock} is dominated by the scale height below $r_S$ where $L_{\rm max}$ and $c_s$ become small and the integrand becomes large.

At the shock radius $r_S$, the waves steepen into a train of weak shocks. In our implementation below, we assume the waves instantly behave as ``mature shocks" (see discussion in \citealt{ro:17}),  in reality, this occurs at a slightly larger damping radius $r_h$. \cite{ro:17} estimate the maximum shock maturation radius (Equation 29) by approximating the post-shock state to match the wave peak. A train of shocks or an ``N'' wave, which interests us here, propagates slower due to the inward-traveling rarified wave. While this is a seemingly benign point as weak shocks are practically sonic, the relative difference of these speeds gives rise to interesting results. From this knowledge, we derive a conservative bound on the maturation condition
\beq
1 \lesssim \int_{t_S}^{t_h} \frac{4\pi}{\pi-2}\frac{\gamma+1}{2}\sqrt{\frac{L_{\mathrm{wave}}}{L_{\mathrm{max}}}}\omega dt \lesssim 2. 
\eeq
That is, as it is maturing, the wave propagates a distance through the star
\beq
r_h - r_S \sim \frac{c_s}{\omega} \bigg\langle \frac{2\pi(\gamma+1)}{\pi-2}\sqrt{\frac{L_{\mathrm{wave}}}{L_{\mathrm{max}}}} \bigg\rangle^{-1} \, ,
\eeq
where the bracket is the time average along the wave. This distance can be large if the shock forms when $L_{\rm wave} \ll L_{\rm max}$. In our models, however, the waves shock where $L_{\rm wave} \sim L_{\rm max}$, the maturation length is shorter than a scale height, and our neglect of this process is warranted.

Above the point of shock maturation, the shock amplitude changes due to competing effects: energy dissipation at the shock front that decreases the shock amplitude, and shock steepening as the waves propagate into lower density material.  \cite{mihalas:84} assume each segment of an N-wave is \textit{dynamically} similar to that of a single shock with twice the amplitude. This is both incorrect, as discussed previously, and an unnecessary assumption when they model the energy dissipation rate for a shock train. The dissipation rate can be written in terms of the shock strength $x=(P_2-P_1)/P_1$ (note a missing factor of $\gamma+1$ from (56.56) of \citealt{mihalas:84}):
\beq
\label{weakshock}
\frac{d E}{dm} = \frac{(\gamma+1) P}{12\gamma^2 \rho} x^3 \,
\eeq
where $P_1$ and $P_2$ and the pre- and post-shocked pressures and the background indicated with no subscript. Assuming the shocks remain self-similar in shape, which is valid for weak, planar shocks \citep{landau:59,ulmschneider:70}, then the wave strength is a function of the wave energy
\beq
x \simeq \frac{2(P_2-P)}{P} = 2 \gamma\sqrt{\frac{L_{\mathrm{wave}}}{L_{\mathrm{max}}}},
\label{mdef}
\eeq
which is no longer constant due to shock dissipation. Additionally, the change in energy flux per unit mass traversed by the N-wave packet is simply equation \ref{weakshock} multiplied by the linear wave frequency $\omega/(2 \pi)$. The reduction in wave luminosity per unit mass, due to weak shock dissipation is thus
\beq
\label{weakshock2}
\frac{d L_{\rm wave}}{dm} = \frac{\gamma +1}{3 \pi} \omega c_s^2 \bigg( \frac{ L_{\rm wave} }{ L_{\rm max} }\bigg)^{3/2} \, .
\eeq
This expression (up to factors of unity) has been employed for the solar coronal heating problem (see \citealt{ulmschneider:70} for an extensive list).

Equation \ref{weakshock2} can be recast as an effective damping mass, $M_{\rm damp} = L_{\rm wave} (dL_{\rm wave}/dm)^{-1}$, yielding
\beq
\label{Mdampshock}
M_{\rm damp,shock} = \frac{3 \pi}{\gamma + 1} \frac{L_{\rm max}}{\omega c_s^2} \bigg( \frac{L_{\rm max}}{L_{\rm wave}} \bigg)^{1/2} \, .
\eeq
This can be compared with the damping mass due to radiative diffusion,
\beq
\label{Mdamprad}
M_{\rm damp,rad} = \frac{2 L_{\rm max}}{\omega^2 K} \, ,
\eeq
with $K$ the local thermal diffusivity (see \citealt{fuller:17}). In our MESA evolutions, we include both forms of damping, such that the energy deposited per unit mass per unit time in each cell is
\beq
\label{Mdamptot}
\epsilon_{\rm wave} = \frac{dL_{\rm wave}}{dm} = \frac{L_{\rm wave}}{M_{\rm damp,shock}} + \frac{L_{\rm wave}}{M_{\rm damp,rad}}
\eeq

\begin{figure}
\begin{center}
\includegraphics[scale=0.33]{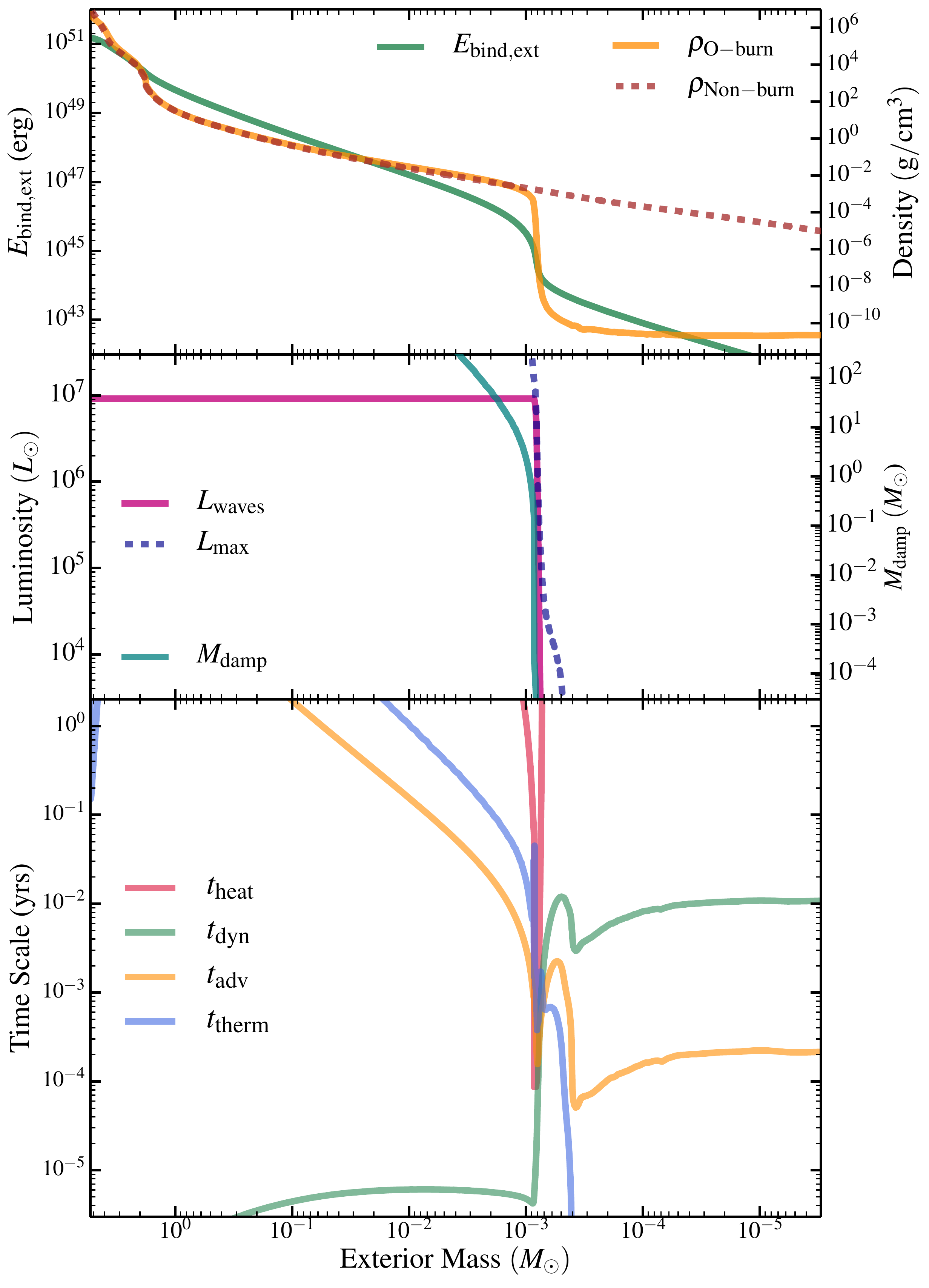}
\end{center} 
\caption{ \label{fig:WaveDamp15Msun_stripped} 
{\bf Top:} Binding energy integrated inward from the surface of our hydrogen-free model, as a function of exterior mass $M_{\rm ext}$, just after carbon burning. The right axis shows the stellar density profile just after carbon shell burning and during core oxygen burning. {\bf Middle:} Wave energy flux $L_{\rm waves}(r)$ during oxygen burning, the maximum possible linear wave flux $L_{\rm max}$, and the damping mass $M_{\rm damp}$ through which the waves must propagate to be attenuated (equation \ref{Mdamptot3}). $M_{\rm damp}$ drops sharply outside the core, causing waves to damp at $M_{\rm ext} \simeq 9 \times 10^{-4} \, M_\odot$ such that $L_{\rm waves}$ drops where wave energy is converted to heat. {\bf Bottom:} Wave heating timescale, $t_{\rm heat}$, along with the local dynamical timescale $t_{\rm dyn}$, advective timescale by the outflow $t_{\rm adv}$, and local thermal timescale $t_{\rm therm}$.  }
\end{figure}

\subsubsection{Accounting for Background Flows}
\label{flows}

Near and above the wave heating region, material is accelerated upward by the pressure gradient resulting from wave heat. Consequently, the waves may propagate into regions where the background flow velocities are large or even supersonic. Hence, we must include the effects of these background flows on acoustic wave propagation.

For acoustic waves in the WKB limit, it can be shown that radial background flows simply alter the wave dispersion relation to 
\beq
\label{kflow}
k_r \simeq \frac{\pm \omega}{c_s \pm v_r} \, ,
\eeq
where $\omega$ is the inertial frame wave frequency, $v_r$ is the background flow velocity, and the plus and minus refer to outwardly and inwardly propagating waves, respectively. For the outwardly propagating waves of interest, we can see that their wave numbers are decreased in the presence of an outward background flow, which is essentially just a Doppler shift. The outward flows generally lengthen the wave damping region due to the reduction in wavenumber. For supersonic inward flows there is a critical point where $v_r = -c_s$, at which point the wavenumber diverges such that the waves will be absorbed. The wave group velocity (in the inertial frame) is simply $v_g = c_s + v_r$. 

In the presence of flows, the wave action (rather than wave energy flux) is conserved in the absence of damping. The wave action is 
\beq
\label{Lac}
L_{\rm ac} = 2 \pi r^2 \rho c_s u^2 (1 + v_r/c_s) \simeq {\rm constant} \, .
\eeq
Incorporating the effects of background flows, the damping masses become
\beq
\label{Mdampshock2}
M_{\rm damp,shock} = \frac{3 \pi}{\gamma + 1} \frac{L_{\rm max}}{\omega c_s^2} \bigg( \frac{L_{\rm max} (1 + v_r/c_s)^5}{L_{\rm ac}} \bigg)^{1/2} \, .
\eeq
and
\beq
\label{Mdamprad2}
M_{\rm damp,rad} = \frac{2 L_{\rm max} (1 +v_r/c_s)^2}{\omega^2 K} \, ,
\eeq
with net wave dissipation rate
\beq
\label{epswave}
\epsilon_{\rm wave} = \frac{L_{\rm ac}}{M_{\rm damp,shock}} + \frac{L_{\rm ac}}{M_{\rm damp,rad}} \, .
\eeq
The effective damping mass is 
\beq
\label{Mdamptot3}
M_{\rm damp} = \bigg[ M_{\rm damp,shock}^{-1} + M_{\rm damp,rad}^{-1} \bigg]^{-1} \, .
\eeq
Equation \ref{epswave} is the wave heating rate implemented in our models.

\subsection{Summary of Model Physics}

Here we summarize the implementation wave heating into our models. We begin adding wave heat and using hydrodynamics just before core carbon burning as described in Section \ref{mods}. At each timestep, we perform the following steps:

1. Calculate the IGW luminosity $L_{\rm wave}$ and characteristic wave frequency generated by core  convection, as described in Section \ref{gen}. 

2. Calculate the fraction of wave energy, $f_{\rm esc}$ that propagates into the envelope as described in Section \ref{prop}.

3. Calculate the envelope wave heating rate at each cell in the envelope (above the point where IGW transform to acoustic waves) as described in Section \ref{shocks}.

4. Add the wave heat in each cell of the stellar model. This is the only effect of the waves that we  add to the models.

5. Evolve the model to the next time step.

\section{Wave-driven Outflows}

It is important to discuss the physics of the super-Eddington outflows that arise in our stellar models. More detailed discussion of such outflows are presented in \cite{quataert:16,shen:16,owocki:17}, here we discuss those results in the context of super-Eddington outbursts driven by wave heating. The important characteristics of the wave heating in our models is that it occurs close enough to the stellar surface that the wave energy deposition exceeds the binding energy of overlying material, yet deep enough within the star that overlying material is very optically thick. Under these conditions, the wave heat will drive a dense, super-Eddington wind as discussed by \cite{quataert:16}.

\autoref{fig:WaveDamp15Msun_stripped} shows profiles of density, binding energy, wave energy flux, and various timescales within the star. Timescales are defined in \cite{fuller:17}, with the addition of an advective timescale
\beq
\label{tadv}
t_{\rm adv} = \frac{H}{v} \,
\eeq
where $H=P/(\rho g)$ is the scale height and $v$ is the local outflow velocity. In contrast to hydrogen-rich stars, we find in our hydrogen-free model that waves typically damp near the stellar surface where the overlying mass is of order $M_{\rm ext} \sim 10^{-3} \, M_\odot$, and the overlying binding energy is only $E_{\rm ext} \sim 10^{45} \, {\rm erg}$. However, the optical depth is still quite large, of order $\tau \sim 10^4$, so wave heat cannot immediately diffuse outward. At the wave heating location, $t_{\rm heat} < t_{\rm therm}$, so material heats up faster than it cools, and its pressure increases. The heated material drives an outflow that is accelerated by the gradient in radiation pressure between the hot wave heating region and the cooler expanding material above. In equilibrium, the outflow is accelerated such that $t_{\rm heat} \sim t_{\rm adv}$ in the heating region, i.e., wave heat is advected up by the outflow at the same rate that it is deposited. The outflow sonic point occurs where $t_{\rm adv}=t_{\rm dyn}$. For wave heating rates of $\sim \! 10^7 \, L_\odot$ sustained for periods of $\sim \! 1 \, {\rm yr}$, the waves deposit more than enough energy to unbind the overlying envelope and the heating is highly super-Eddington. Hence, the waves drive an optically thick super-Eddington wind that is accelerated in and above the wave heating region.

\begin{figure}
\begin{center}
\includegraphics[scale=0.34]{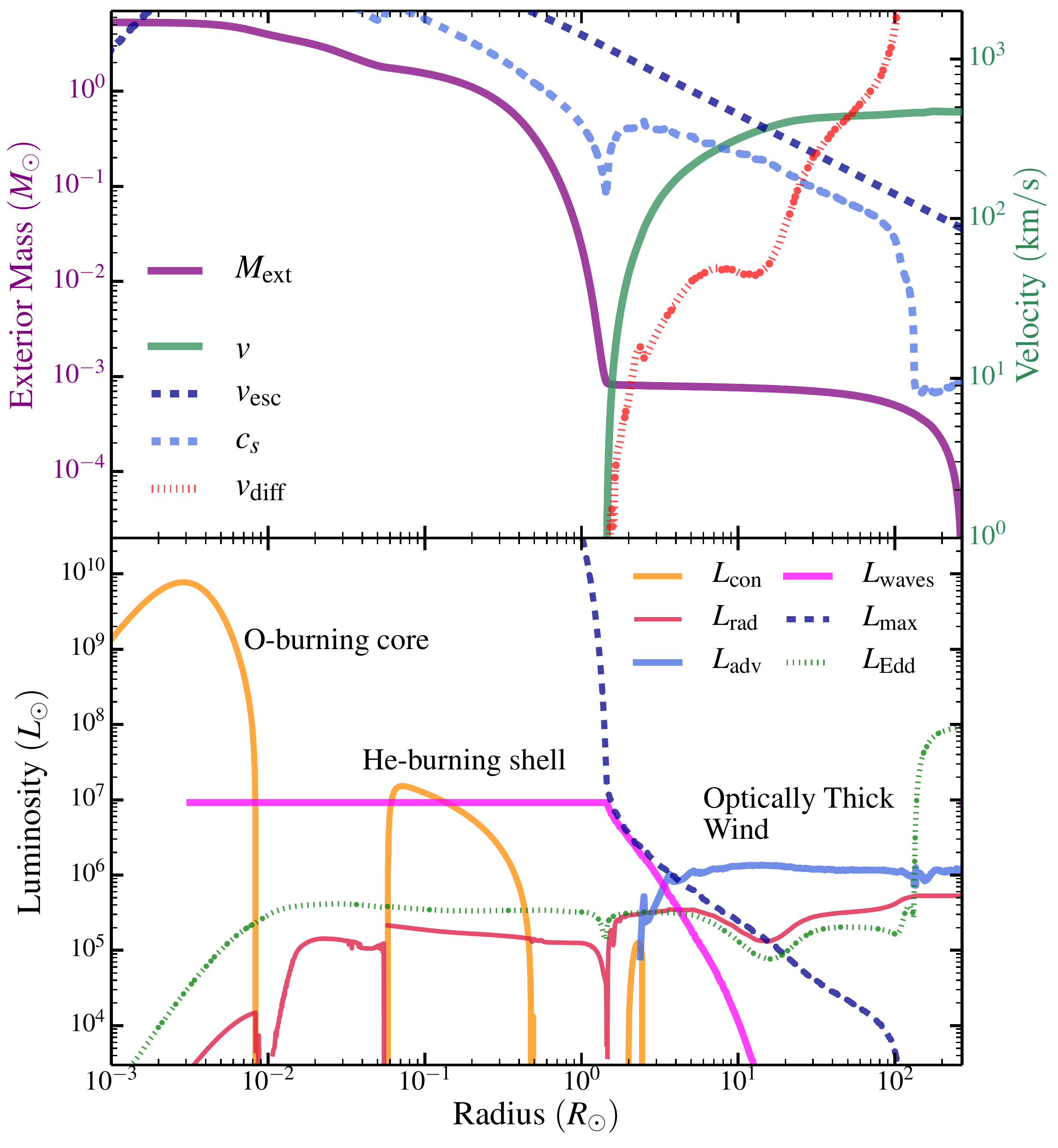}
\end{center} 
\caption{ \label{fig:Structure15Msun_stripped} 
{\bf Top:} Exterior mass $M_{\rm ext}$ and outflow velocity $v$ as a function of radial coordinate in our hydrogen-free model during core oxygen burning. We also show the local sound speed $c_s$, escape speed $v_{\rm esc}$, and photon diffusion speed $v_{\rm diff}$. {\bf Bottom:} Energy flux carried by convection, $L_{\rm con}$, radiation, $L_{\rm rad}$, advection by a wind, $L_{\rm adv}$, waves, $L_{\rm waves}$, maximum possible linear wave flux $L_{\rm max}$, and local Eddington luminosity $L_{\rm edd}$. The wave energy flux decreases at the damping region ($r \approx 2 \, R_\odot$, $M_{\rm ext} \approx 9 \times 10^{-4} \, M_\odot$). Most of this energy is used to unbind mass, with a small fraction emerging in the wind's advective energy flux and radiative flux. }
\end{figure}

These sorts of very optically thick, very super-Eddington winds behave much differently from optically thin or near-Eddington winds (see discussion in \citealt{owocki:17}) such as line-driven winds from stars or star-forming regions. A key difference is that photons are trapped within the outflowing gas, both in regions near the sonic point $r_S$, and the point where the outflow exceeds the escape speed $r_{\rm esc}$. The gas can be treated as a single fluid and is not subject to two-fluid instabilities. The photons' momentum is unimportant for expelling material, instead, it is the gradient in radiation pressure that drives the outflow. Figure \ref{fig:Structure15Msun_stripped} shows that the photon diffusion speed,
\beq
\label{vdiff}
v_{\rm diff} = F_{\rm rad}/a T^4 \, ,
\eeq
is much smaller than the wind speed at both $r_S$ and $r_{\rm esc}$. This means the wind behaves nearly adiabatically as discussed in \cite{quataert:16}. Requiring that $v_{\rm diff} < v$ is equivalent to the statement that the optical depth satisfies $\tau > c/v$ at $r_S$ and $r_{\rm esc}$. In this limit, the outflow dynamics simplify, radiative diffusion is a good approximation, and multi-dimensional radiative instabilities are unlikely to be important in the wind-launching region. Such instabilities are expected to grow on a timescale $t \sim r/v_{\rm diff}$, which is much larger than the outflow timescale $r/v$, so our estimates of wind speeds, mass loss rates, and kinetic energy flux are realistic. Multi-dimensional hydrodynamical instabilities (e.g., convection) are also unlikely to be important as shown by the simulations presented in \cite{quataert:16}.

\autoref{fig:Structure15Msun_stripped} shows that during intense wave heating episodes such as during core oxygen burning, most of of the wave energy is converted into gravitational potential energy. During oxygen burning, roughly 85\% of the wave energy is used to lift mass out of the gravitational well of the star and expel it to infinity. Rougly 10\% of the energy escapes to infinity in the advective energy flux of the wind, which we define as 
\beq
\label{Ladv}
L_{\rm adv} = 4 \pi r^2 \rho v_r \big( c_s^2 + v_r^2/2 \big) \, .
\eeq
At infinity, the wind is highly supersonic and the advective flux is essentially the kinetic energy flux of the wind. Note that at $r_S$ and $r_{\rm esc}$, the wave kinetic energy flux is well above the Eddington luminosity $L_{\rm Edd}$. This again proves the outflow is nearly adiabatic \citep{quataert:16}, such that photon diffusion is negligible in regions where the outflow is launched. Finally, roughly 5\% of the wave energy is radiated in photons. The emergent photon luminosity is super-Eddington (for electron scattering opacity), but only by a factor of a few. Our computed radiative luminosity (based on the diffusion approximation) is also likely to be realistic because the model luminosity is nearly constant where $v_{\rm diff} \sim v$.  Near the photosphere, of course, the adiabatic approximation breaks down, and it may be possible that our 1D models incorrectly predict the effective temperature of the outflow. However, the photosphere occurs near the drop in opacity caused by H or He recombination ($r\sim 150 \, R_\odot$ in Figure \ref{fig:Structure15Msun_stripped}), like it does in many SNe. Therefore, the effective temperature is mostly determined by the recombination temperature, and not by the dynamics of the outflow, and our estimates are probably reasonable.

It is instructive to compare our numerical results on super-Eddington mass loss with analytic expectations. During oxygen burning in our hydrogen-free model, there is a nearly steady-state super-Eddington wind. Following the formalism of \cite{quataert:16}, our model wind is driven by a heating rate $\dot{E} \! \sim \! 10^7 \, L_\odot$, deposited at a radius $r_h \! \sim \! 2 \, R_\odot$, with overlying mass $M_{\rm env} \! \sim \! 10^{-3} \, M_\odot$ and interior mass $M \! \sim \! 5 \, M_\odot$. Our model thus has $v_{\rm crit} \! \sim \! 300 \, {\rm km}/{\rm s}$, $v_{\rm esc}(r_h) \! \sim \! 1000 \, {\rm km}/{\rm s}$, and $f \! \sim \! 1$, and lies in the mass-loaded Regime 2 of \cite{quataert:16}, where most of the wave energy is used to lift material out of the star's gravitational potential well, as discussed in Section \ref{Ib}. The analytics of \cite{quataert:16} predict a mass loss rate $\dot{M} \sim 0.1 \, M_\odot/{\rm yr}$, terminal wind speed $v_\infty \sim 300 \, {\rm km}/{\rm s}$, and wind kinetic energy flux $\dot{E}_w/\dot{E} \sim 0.06$, which are very close to the values discussed in Section \ref{Ib} and shown in Figures \ref{fig:Structure15Msun_stripped} and \ref{fig:Mdotv15Msun_stripped}. Additionally, the Eddington luminosity of this model is $L_{\rm Edd} \! \sim \! 2 \times 10^5 \, L_\odot$, and the predicted photon luminosity of the wind is $L_{\rm rad} \! \sim \! 7 \times 10^5 \, L_\odot$, similar to that shown in Figure \ref{fig:Luminosity15Msun_stripped}. In the language of \cite{owocki:17}, our model is somewhat photon tired, with $m \approx 0.8$, but still can maintain a strong stable outflow because of the greatly super-Eddington heating rate such that $m \Gamma_o \gg 1$. For hydrogen-rich models like the red supergiant models in \cite{fuller:17}, a steady state wind is not produced, and these analytic predictions are not applicable. The hydrogen-poor model of Section \ref{IIb} is an intermediate case, nearly reaching a steady state at the end of core oxygen burning, but not early on when the hydrogen envelope is still being ejected.

As overlying material is blown off by the wind, it is replaced by upwelling material from deeper in the star. Hence, the wave heating digs deeper into the star, moving to smaller mass coordinates as material is blown off in the wind. The outer core (above the wave generation region, but below the wave heating region) expands adiabatically in response to this mass loss. The inner core evolution is essentially unaffected since it is determined primarily by the mass of the carbon core, which is not affected by the wave-induced mass loss. 

%{\bf A possible caveat with our implementation of wave heating is that, as shown by \autoref{fig:Structure15Msun_stripped}, the waves damp where $L_{\rm waves}}

\section{Effect on Pre-SN Evolution}

\subsection{Type Ib progenitor}
\label{Ib}

\begin{figure}
\begin{center}
\includegraphics[scale=0.35]{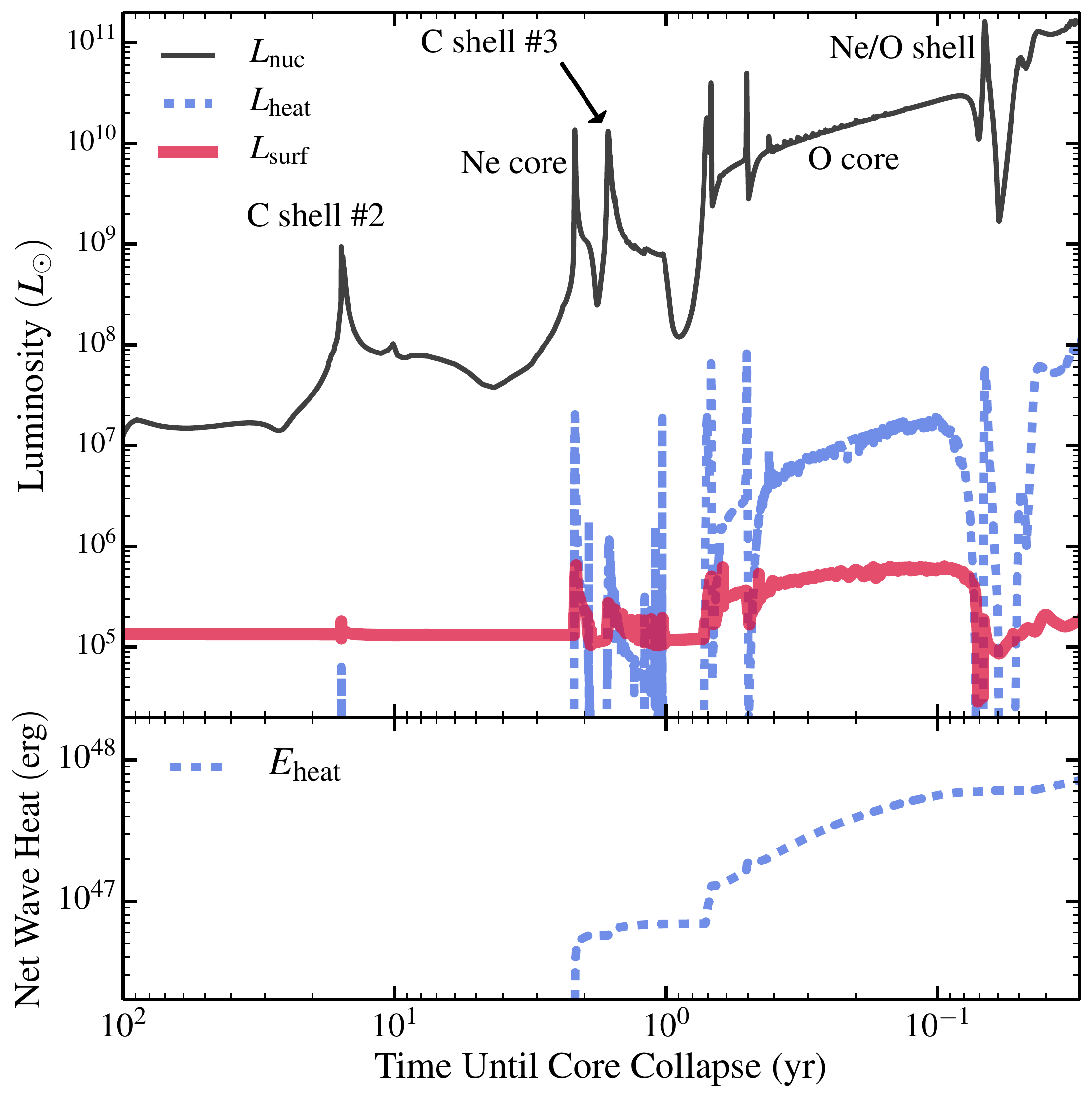}
\end{center} 
\caption{ \label{fig:Luminosity15Msun_stripped} 
{\bf Top:} Luminosities of our hydrogen-free model as a function of time before core-collapse. Burning phases are labeled next to the nuclear energy generation rate $L_{\rm nuc}$. We also plot the envelope wave heating rate $L_{\rm heat}$, and photospheric radiative luminosity $L_{\rm surf}$. {\bf Bottom:} Net wave energy deposited in the envelope, i.e., the integral of $L_{\rm heat}$ in top panel.}
\end{figure}

Here we examine the observable surface properties of our model containing no hydrogen. \autoref{fig:Luminosity15Msun_stripped} shows the core nuclear energy generation rate, the wave heating rate of the stellar envelope, and the surface luminosity of our stellar model. Before neon burning, the wave heating rate remains below the star's radiative luminosity, and the effect of wave heating is minor. When neon burning begins $\sim \! 2$ years before explosion, wave heating rates exceed $10^6 \, L_\odot$, and they become both super-Eddington and larger than the background energy flux (i.e., the stellar luminosity without including wave energy transport). The wave heat drives a dense wind from the surface of the star, such that the photosphere lies within this wind, typically above both the wind sonic point, and above the point where the wind velocity exceeds the escape velocity.

\autoref{fig:Luminosity15Msun_stripped} shows that the photospheric luminosity is much smaller than the wave heating rate during these late burning phases. Most of the wave energy is used to unbind mass, and most of the remaining energy is put into the kinetic energy of the wind. Our models predict that the radiated luminosity remains below $\sim 10^6 \, L_\odot$. Nevertheless, stars undergoing wave-driven outbursts may increase their luminosity by a factor of $\sim 5$, which could be detectable by ground-based surveys such as the ongoing LBT survey \citep{kochanek:08,adams:17}, or the future LSST survey. Outburst luminosities at visual wavelengths will likely be even larger in relative brightness, due to bolometric corrections and additional shock heating from shell-shell collisions or interaction with a companion star (see \autoref{disc}).

\begin{figure}
\begin{center}
\includegraphics[scale=0.34]{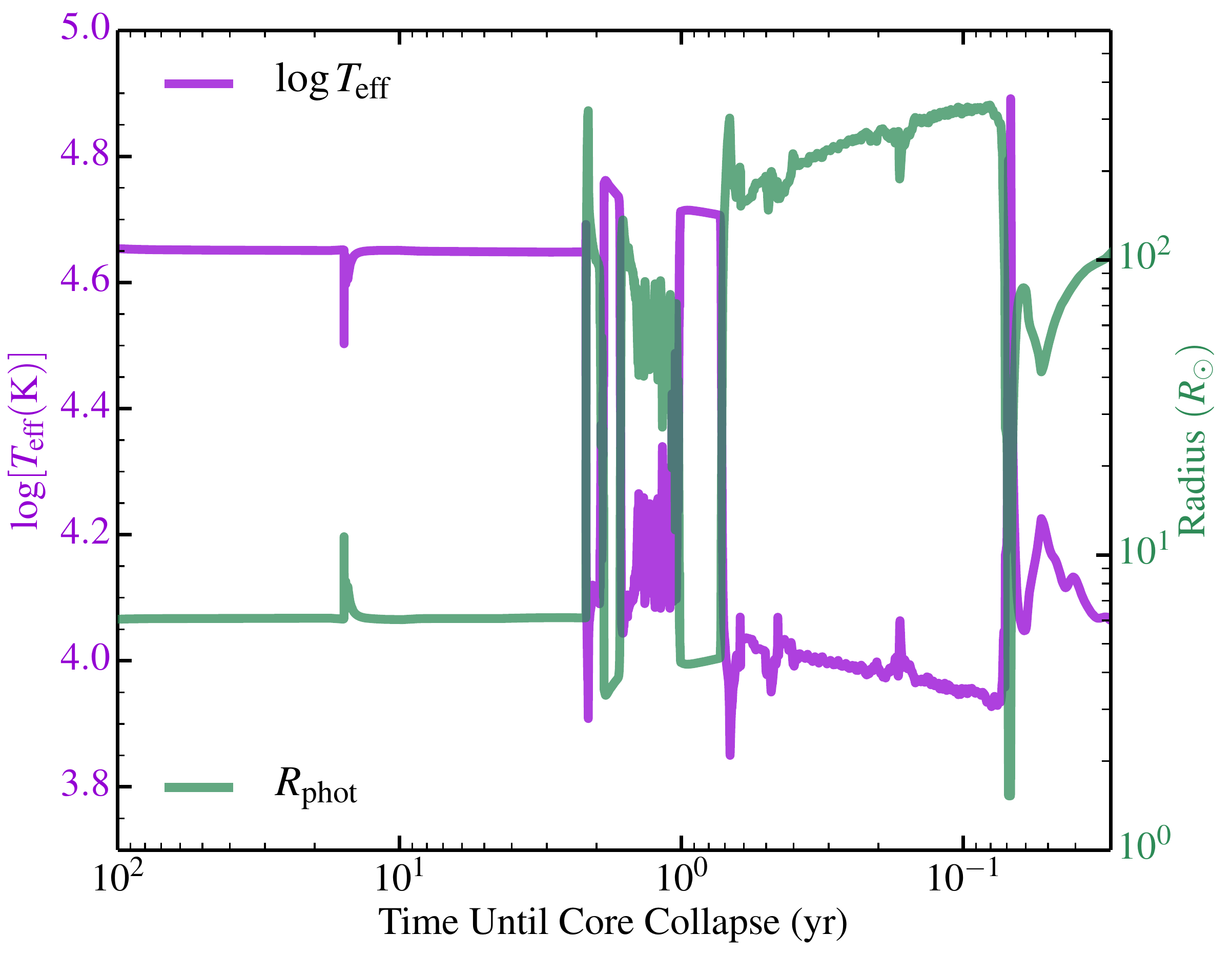}
\end{center} 
\caption{ \label{fig:TeffR15Msun_stripped} 
Effective temperature and photospheric radius of our hydrogen-free model as a function of time before core-collapse, $t_{\rm col}$. Before the onset of neon burning at $t_{\rm col}$, the star is a hot and compact Wolf-Rayet-like star in its ``low hard" state. During strong wave heating episodes, the  photosphere moves out into the optically thick wind, greatly increasing its radius and decreasing its temperature, creating a ``high soft" state. }
\end{figure}

The wave-driven outbursts have a strong effect on the photospheric temperature and radius (\autoref{fig:TeffR15Msun_stripped}). During quiescence, our model is essentially a low-luminosity Wolf-Rayet star, with an effective temperature of $T_{\rm eff} \sim 5 \times 10^4 \, {\rm K}$, and photospheric radius of $R \sim 5 \, R_\odot$. However, during outbursts, the photosphere moves far out into the optically thick wind driven by the wave heating. The photosphere is actually much {\it cooler} during outburst, with $T_{\rm eff} \sim 10^4 \, {\rm K}$. However, the change in photospheric radius more than compensates for the decrease in temperature, increasing by a factor of $\sim \! 50$ to radii of $\sim 250 \, R_\odot$. As discussed above, the bolometric luminosity increases by a factor of $\sim \! 5$, but the V-band luminosity will increase by a substantially larger factor due to the changing temperature, with the Wien peak moving from $\sim \! 60 \, {\rm nm}$ during quiescence to $\sim 300 \, {\rm nm}$ during outburst. Generally, the star is in a ``low hard" state during quiescence, and a ``high soft" state during outburst.

\begin{figure}
\begin{center}
\includegraphics[scale=0.33]{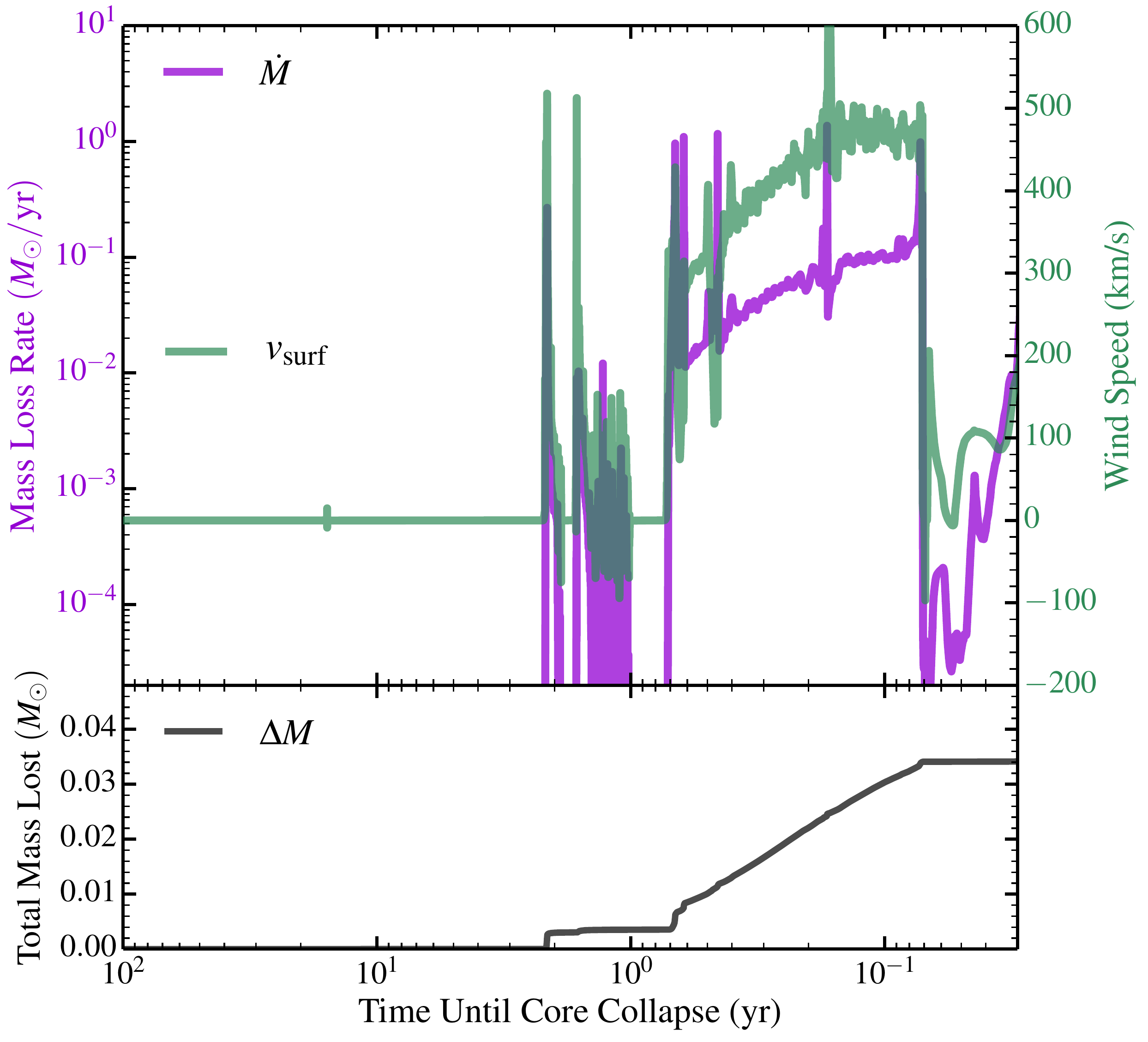}
\end{center} 
\caption{ \label{fig:Mdotv15Msun_stripped} 
{\bf Top:} Mass loss rate $\dot{M}$ and photospheric velocity $v_{\rm surf}$ of our hydrogen-free model as a function of time until core collapse. During outbursts, the photospheric velocity is nearly equal to the wind terminal velocity. {\bf Bottom:} Integrated mass lost due to wave-driven outbursts.}
\end{figure}

The mass loss rate becomes very large during wave-driven outbursts (\autoref{fig:Mdotv15Msun_stripped}). During neon and oxygen burning, we find typical mass loss rates of $10^{-2}-10^{-1} \, M_\odot/{\rm yr}$, a few orders of magnitude larger than expected from line-driven winds in the absence of wave heating. We find typical terminal velocities of $\sim \! 400 \, {\rm km}/{\rm s}$, less than half the escape velocity $\sim \! 1000 \, {\rm km}/{\rm s}$ at the wave heating region. As discussed above, the fact that $v_{\rm term} \! \leq \! v_{\rm esc}$ shows that most of the wave energy is used to lift material out of the star's potential, rather than being used to accelerate the wind to large velocities. Thus, in our models, the wind is somewhat mass-loaded. The total mass lost in the wave-driven wind is $\sim 0.035 \, M_\odot$, most of which is lost during oxygen burning during the final year of the star's life.

\begin{figure}
\begin{center}
\includegraphics[scale=0.37]{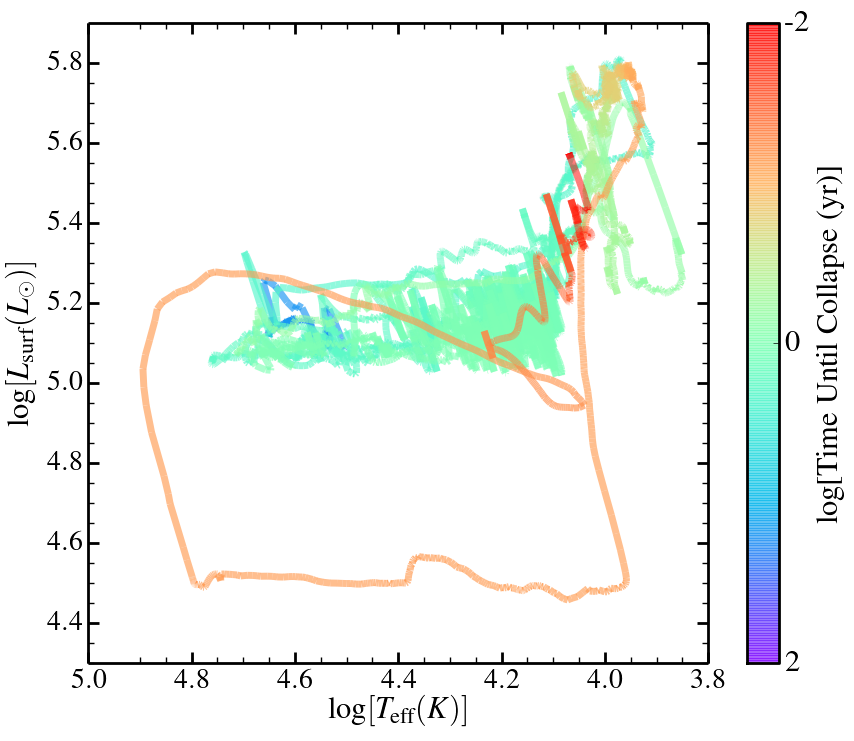}
\end{center} 
\caption{ \label{fig:HR15Msun_stripped} 
Evolution of our hydrogen-free model in the HR diagram during its final years before core-collapse. It spends most of its time in a ``low hard" quiescent state at $\log T \approx 4.65$, $\log L \approx 5.1$, or in a ``high soft" outbursting state at $\log T \approx 4.0$, $\log L \approx 5.7$. Erratic evolution between these states occurs due to stagnating outflows and fallback shocks that are subject to non-radial instabilities that could alter the results.}
\end{figure}

The evolution of the star on a HR diagram is shown in \autoref{fig:HR15Msun_stripped}. Before wave heating, the star has $\log(T_{\rm eff}/K) \approx 4.7$ and $\log(L/L_\odot) \approx 5.1$ while it is in its low hard state. During wave-driven outbursts, the star moves to its high soft state with $\log(T_{\rm eff}/K) \approx 4.0$ and $\log(L/L_\odot) \approx 5.7$. When burning phases turn off and on, and during low-amplitude outbursts (e.g., during carbon shell burning at $\sim 1.2 \, {\rm yr}$ before explosion), the star is highly variable, moving erratically in the HR diagram. In the models, this variability typically occurs when the wave heating rate is only moderately super-Eddington, and the wave-driven outflows are not efficiently driven above the escape velocity. The outflow becomes unstable, frequently stagnating and forming fallback shocks that make the photospheric luminosity highly variable. We do not trust our results during these phases, as these stagnating outflows are susceptible to multi-dimensional instabilities that will alter the dynamics and luminosity \citep{owocki:15}. 

The excursion to low luminosity in \autoref{fig:HR15Msun_stripped} (i.e., ``the legs of the horse") occurs during the very brief quiescent interval between oxygen core burning and neon shell burning. At this time, wave heating ceases and the optically thick wind disappears, exposing the small and hot Wolf-Rayet-like core. While this feature is prominent in \autoref{fig:HR15Msun_stripped}, it is only a very small segment of the total light curve (corresponding to the dip at $\sim \! 0.07 \, {\rm yr}$ in \autoref{fig:Luminosity15Msun_stripped}), and it may be difficult to observe in practice.

\subsection{Type IIb progenitor}
\label{IIb}

\begin{figure}
\begin{center}
\includegraphics[scale=0.35]{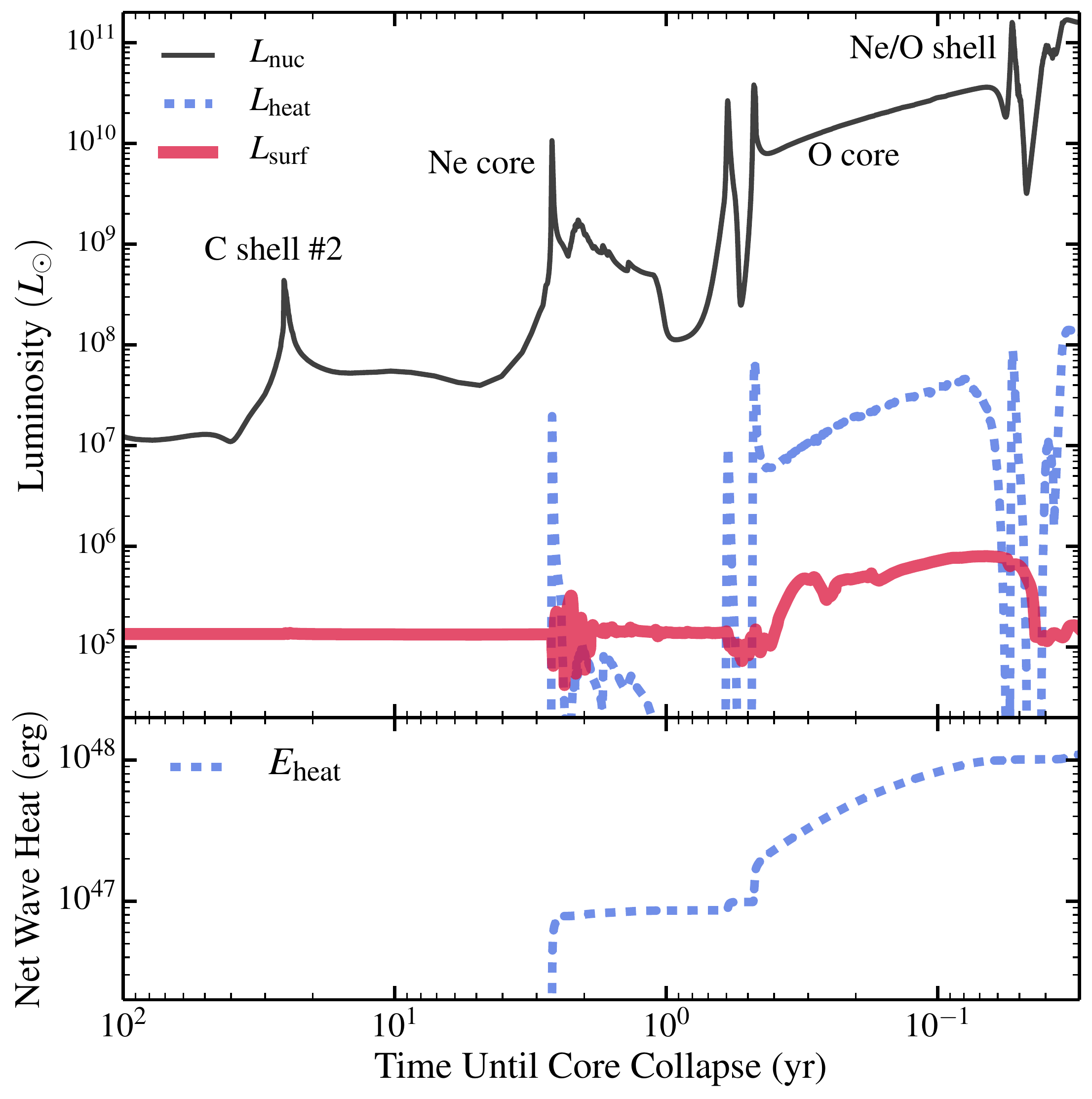}
\end{center} 
\caption{ \label{fig:Luminosity15Msun_stripped_003} 
Same as Figure \ref{fig:Luminosity15Msun_stripped} for our hydrogen-poor model.}
\end{figure}

The model containing a small amount of hydrogen undergoes similar but distinctly different evolution from the hydrogen-free model. \autoref{fig:Luminosity15Msun_stripped_003} shows the wave heating rates and surface luminosity of our model with hydrogen. We note that the core evolution, wave heating rate, and integrated wave heat is similar but not identical to the hydrogen-free model. The difference reflects the sensitivity of core evolution to its structure, in addition to some dependence on the spatial/temporal numerical resolution of the models. Our model with hydrogen has a slightly larger helium core mass and a slightly shorter and more energetic oxygen burning phase, leading to a slightly larger wave heating rate and net wave heat deposited in the envelope.

Compared to the hydrogen-free model, the model with hydrogen has a smaller surface luminosity during the core neon burning. The hydrogen envelope helps prevents wave heat from diffusing outward, and instead the wave energy is used to eject more mass from the hydrogen envelope. By the time of oxygen burning, however, the hydrogen envelope has been mostly ejected, and the surface luminosity is similar to the model hydrogen-free model.

\begin{figure}
\begin{center}
\includegraphics[scale=0.34]{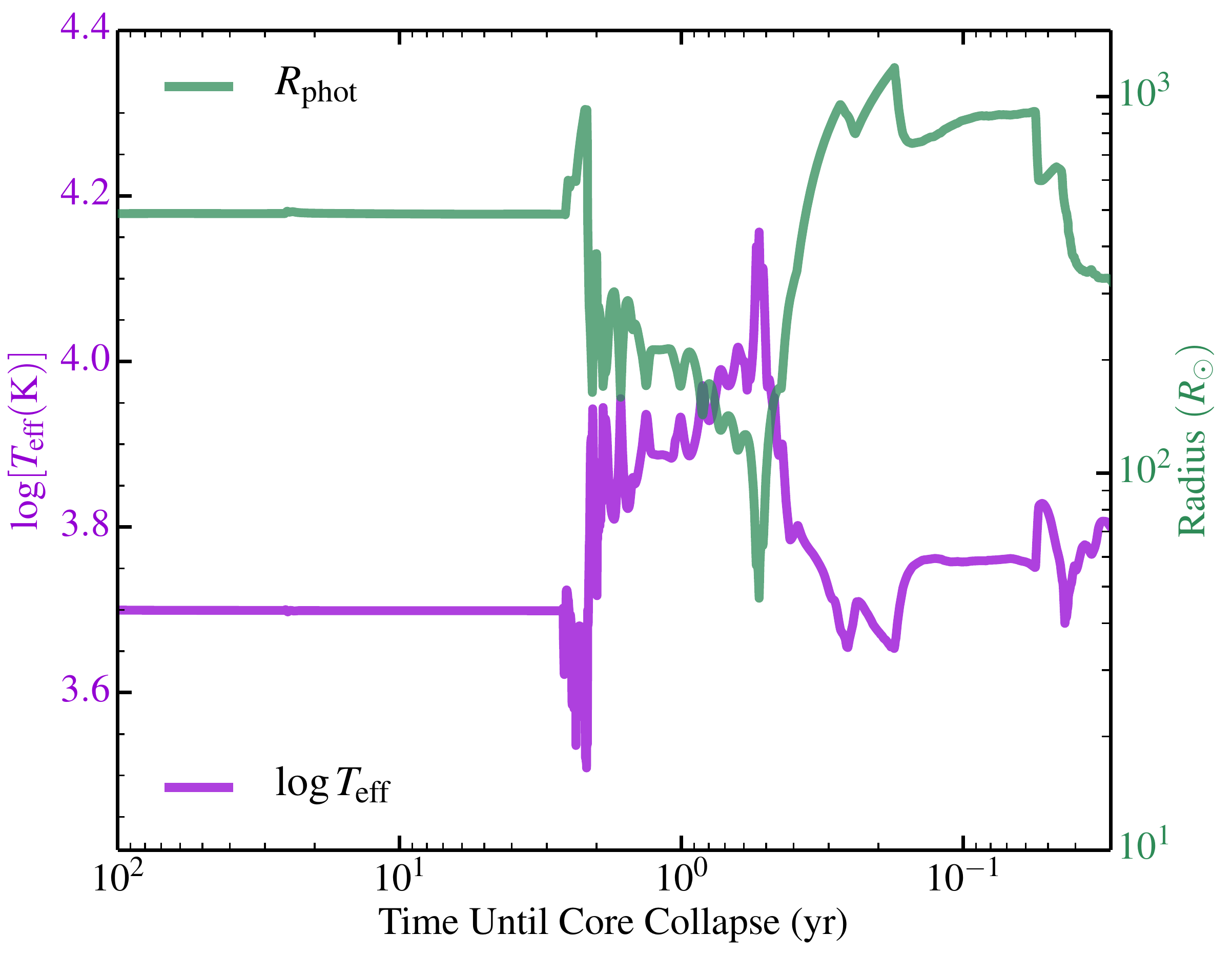}
\end{center} 
\caption{ \label{fig:TeffR15Msun_stripped_003} 
Same as Figure \ref{fig:TeffR15Msun_stripped} for our hydrogen-poor model. The presence of hydrogen keeps this star cooler and more extended, though it becomes slightly hotter and more inflated during outburst.}
\end{figure}

Despite similar energetics, the photospheric radius and temperature of the model with hydrogen are significantly different (\autoref{fig:TeffR15Msun_stripped_003}). In general, the presence of hydrogen leads to larger opacity per unit mass, and hence larger photospheric radii and cooler effective temperatures than the hydrogen-free model. For instance, even during oxygen burning when most of the hydrogen-rich envelope has already been ejected, a small amount of hydrogen lingers in the remaining helium-rich layers, and the typical photospheric radius is 2-3 times larger in the model with hydrogen. Correspondingly, the effective temperature of the model with hydrogen ranges between $5 \! \times \! 10^{3} \, {\rm K} \! < \! T_{\rm eff} \! < \! 12 \! \times \! 10^{3} \, {\rm K}$, whereas the hydrogen-free model ranges from $9 \! \times \! 10^{3} \, {\rm K} \! < \! T_{\rm eff} \! < \! 50 \! \times \! 10^{3} \, {\rm K}$.

\begin{figure}
\begin{center}
\includegraphics[scale=0.34]{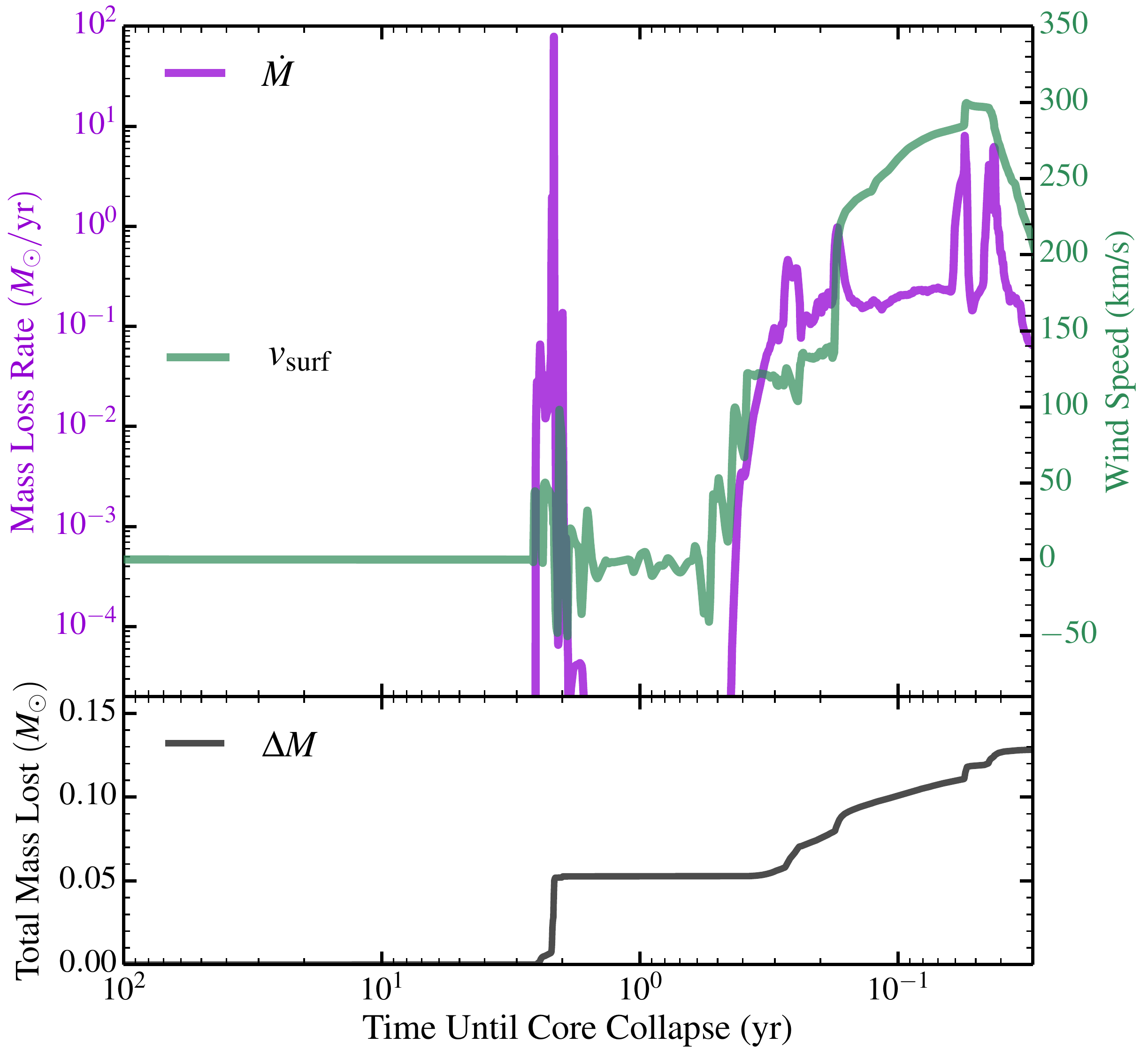}
\end{center} 
\caption{ \label{fig:Mdotv15Msun_stripped_003} 
Same as Figure \ref{fig:Mdotv15Msun_stripped} for our hydrogen-poor model. The entire hydrogen-rich envelope is ejected by wave-driven outbursts.}
\end{figure}

\autoref{fig:Mdotv15Msun_stripped_003} shows that the mass loss rate, and total mass lost, are significantly larger for the model with hydrogen. This is not surprising, as its low density hydrogen envelope extends to much larger radii, and is much more weakly bound than the surface layers of the hydrogen-free model. During core neon burning, essentially all of the extended hydrogen-rich material is ejected, leading to a temporarily extreme mass loss rate of $\dot{M} \! > \! 1 \, M_\odot/{\rm yr}$. After neon burning, the model becomes much smaller in radius during the quiescent phase before oxygen burning (see \autoref{fig:TeffR15Msun_stripped_003} at $\sim \! 1 \, {\rm yr}$ before core-collapse). The surface layers become mostly helium, but they still contain a small fraction of hydrogen, and they are still much more extended than the hydrogen-free model. Hence, the mass loss rate during oxygen burning remains quite large, and the model loses a total of more than $0.1 \, M_\odot$, despite the fact that its total hydrogen content at carbon burning was only $0.03 \, M_\odot$. By the time of core-collapse, essentially all of the hydrogen has been ejected in the super-Eddington wind, and less than $3 \! \times \! 10^{-5} \, M_\odot$ of hydrogen remain.

\begin{figure}
\begin{center}
\includegraphics[scale=0.37]{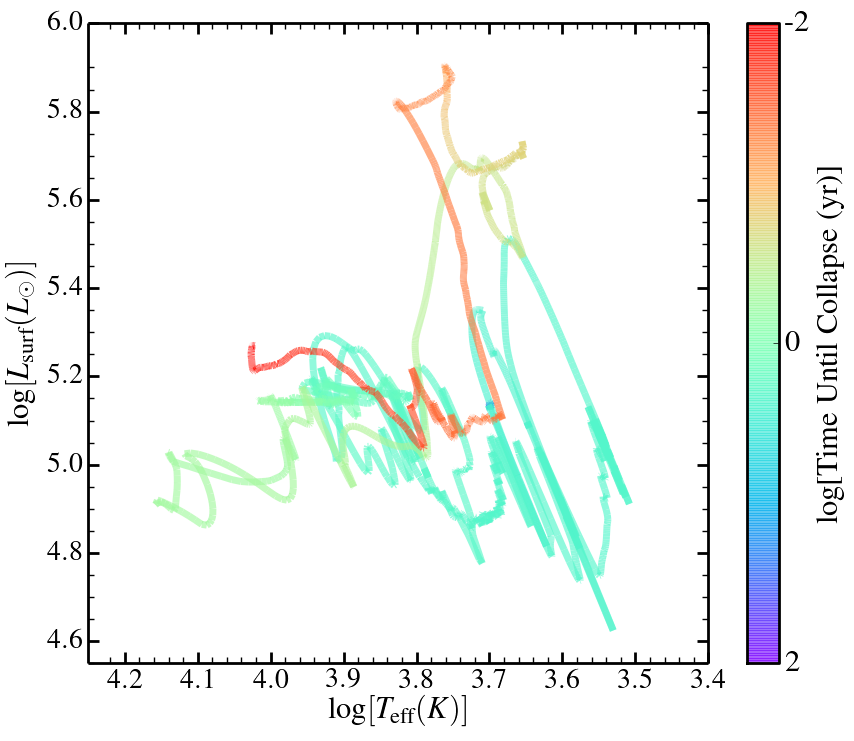}
\end{center} 
\caption{ \label{fig:HR15Msun_stripped_003} 
Same as Figure \ref{fig:HR15Msun_stripped} for our hydrogen-poor model. Temperature changes are smaller than the hydrogen-free model, but changes in luminosity between outburst and quiescence are similar.}
\end{figure}

The HR-diagram evolution of our model with hydrogen is shown in \autoref{fig:HR15Msun_stripped_003}. Similar to the hydrogen-free model, the model with hydrogen exhibits both a ``low hard" state during quiescence and a ``high soft" state during wave-driven outbursts, each with a similar bolometric luminosity to the corresponding state of the hydrogen-free model. However, both of these states have significantly lower effective temperatures in the model with hydrogen. We thus expect both quiescent and outbursting progenitors with hydrogen to be redder (though with similar bolometric luminosity) than hydrogen-free progenitors. Like the hydrogen-free model, the model with hydrogen is highly variable in between outburst and quiescence, although we are not confident in the details of this variability, as discussed in \autoref{Ib}.

\section{Discusssion}
\label{disc}

\subsection{Implications for Subsequent Supernovae}

\subsubsection{Hydrogen-free progenitors}

Wave-driven outbursts from hydrogen-free SN progenitors would significantly affect the subsequent SNe. Our hydrogen free model has a typical mass loss rate in the last months of life of $\sim \! 10^{-1} \, M_\odot/{\rm yr}$, and the total mass lost is $\Delta M \! \sim \! 4 \! \times \! 10^{-2} \, M_\odot$. Wind speeds are $v \! \sim \! 400 \, {\rm km}/{\rm s}$, and strong mass loss begins $\sim 2 \, {\rm yr}$ before core-collapse. We thus expect helium-rich wind material to extend $\sim \! 3 \! \times \! 10^{15} \, {\rm cm}$ away from the progenitor. At larger radii, we predict the CSM density will fall sharply, as wave-driven outbursts are negligible during prior phases of evolution. In these models, wave-driven mass loss continues until $\sim \! 0.07 \, {\rm yr}$ before core-collapse, and we expect the inner boundary of dense CSM to extend inward to $\sim \! 10^{14} \, {\rm cm}$. Hence, for SN ejecta velocities of $\sim \! 10^4 \, {\rm km}/{\rm s}$, we expect CSM interaction to begin early (within the first $\sim \! {\rm day}$ after explosion), and to last for roughly one month.

Because the wave-driven mass loss is not steady, but occurs over multiple distinct outbursts, we do not expect the CSM density profile to have a constant mass loss density profile of $\rho \propto r^{-2}$. There may be multiple dense shells of CSM ejected during different wave-driven outbursts associated with different burning phases (e.g., core neon burning, core oxygen burning, carbon/neon/oxygen shell burning). However, some of these shells may be thick and exhibit a nearly wind-like density profile within their radial extent. In the hydrogen-free model, most of the CSM is ejected at somewhat constant mass loss rate and terminal velocity during core oxygen burning. This material will likely dominate the observed SN-CSM interaction near the SN peak during the first $\sim \! 15 \, {\rm days}$ after explosion. Thus, the early time CSM interaction may appear consistent with a steady wind density profile, while CSM density would drop sharply after $\sim \! 1 \, {\rm month}$. We note these interaction timescales are specific to this progenitor model and could be markedly different for lower/higher mass stars, an issue which we hope to explore in future work.

The characteristics of wave-driven mass loss in our model are broadly consistent with those of observed Ibn SNe. Our model mass loss rates are similar to the mass loss rates of $0.2 \, M_\odot/{\rm yr} \! \lesssim \! \dot{M} \! \lesssim \! 1 \, M_\odot/{\rm yr}$ inferred for SN 2015U \citep{shivvers:16}. The wind speeds are also similar, with our terminal velocities of $\sim \! 500 \, {\rm km}/{\rm s}$ squarely in the range (though perhaps on the low end) of observed narrow P-cygni velocities ($200 \, {\rm km}/{\rm s} \! \lesssim \! v \! \lesssim \! 2000 \, {\rm km}/{\rm s}$) in type Ibn SNe \citep{pastorello:16}. We predict dense CSM that is confined to small distances ($\lesssim \! 10^{16} \, {\rm cm}$) from the progenitor, similar to what has been inferred from the short-lived CSM interaction in most type Ibn SNe \citep{hosseinzadeh:17,moriya:17,shivvers:17}. The confined CSM creates a fast rise, bright peak, and fast decline in the lightcurve. It also causes narrow P-cygni profiles to be observed at early times and disappear at late times after the CSM has been swept up by the SN ejecta, as is frequently observed in type Ibn SNe \citep{hosseinzadeh:17,shivvers:16}. Finally, the confined CSM prevents the observation of late-time radio emission like that frequently seen in type IIn SNe \citep{shivvers:17}.

\subsubsection{Hydrogen-poor progenitors}

Our model containing hydrogen also expels dense CSM that will affect the SN lightcurve and spectra. The most notable difference, of course, is the presence of small amounts of hydrogen, which will likely be observable in SNe spectra, at least at early times. In our model, the wave-driven outbursts expel more than $0.1 \, M_\odot$ of H/He-rich material into the CSM, hence its subsequent SN may be observed as an interacting Ibn/IIn SN similar to the transitional SNe discussed in \cite{pastorello:16} and \cite{hosseinzadeh:17}. We also note that the wind speeds of this model are markedly slower, with $100 \, {\rm km}/{\rm s} \! \lesssim \! v_{\rm wind} \! \lesssim \! 300 \, {\rm km}/{\rm s}$. This appears to be consistent with the slower CSM measured in transitional SNe. 

The slower wind speeds of our model with hydrogen also implies the CSM will be limited to even smaller distances around the progenitor. We find $v_{\rm wind} \! \lesssim \! 100 \, {\rm km}/{\rm s}$ during the neon-burning outburst roughly 2 years before explosion, implying the CSM is confined to radii of $R_{\rm CSM} \! \lesssim \! 6 \! \times \! 10^{14} \, {\rm cm}$. Hence, we would expect the dense CSM to be swept up within $\sim \! 7 \,{\rm days}$ after explosion, and the interaction signatures may only be recognized in SNe with sufficiently early spectra. This may prevent some of these SNe from being classified as transitional Ibn/IIn SNe. They may instead be classified as type IIb SNe if spectra are obtained after the hydrogen-rich material has been swept up, but before the photosphere has receded below hydrogen-rich layers. If only late-time spectra are available, the SN may be classified as a type Ib SN. Hence, these events may be more common than the seemingly rare transitional Ibn/IIn SNe identified in the literature to date.

Wave-driven outbursts may also be able to create transformational SNe like SN 2014C \citep{milisavljevic:15,margutti:16}, which transformed from a type Ib SN to a type IIn SNe after several months. However, the hydrogen-rich material in SN 2014C has somewhat different properties from the ejecta of our model with hydrogen: the observed CSM is estimated to be significantly more massive ($M_{\rm CSM} \! \sim \! 1 \, M_\odot$), it extends to much larger radii ($R_{\rm CSM} \sim 6 \times 10^{16} \, {\rm cm}$), and it must have been ejected longer (at least 20 years) before explosion. Hence, the model in this paper cannot reproduce SN 2014C. However, it might be possible that some lower mass stars, which undergo longer carbon/neon/oxygen burning phases, can create transformational SNe like SN 2014C.

Wave-driven mass loss will also affect the shock breakout (SBO) signal of stripped envelope SNe, potentially causing it to occur within the wave-driven wind. The only SBO detected for a hydrogen-poor SN is the {\it Swift} detection of the SBO of type Ib SN 2008D \citep{soderberg:08}. Interestingly, the duration of the high energy emission is significantly longer than predicted from standard stellar models. \citep{svirski:14} show that the observed SBO of 2008D can be produced by a dense stellar wind, requiring enhanced mass loss in the final weeks before the SN. Wave-driven pre-SN mass loss can potentially produce this sort of density structure, though we note that aspherical SBO \citep{couch:11,matzner:13} is also a possibility.

\subsection{Comparison with Progenitor Observations}
\label{progenitors}

Very few progenitor or outburst detections exist for hydrogen-poor SNe. The most famous example is SN 2006jc, which exhibited a bright outburst ($M_R \sim -14.1$, $L_R \sim 4 \times 10^7$) roughly two years before it exploded as a type Ibn SNe \citep{pastorello:07,foley:07}. This timescale is similar to the wave-driven outburst during neon burning of our model. However, the progenitor outburst of SN 2006jc was much more luminous than the outbursts in our models. From the lack of broad helium lines in the spectrum of SN 2006jc, its progenitor may have been more massive and more helium-poor, which may help create more energetic and luminous wave-driven outbursts. Another possibility is that shell-shell collisions in the wave-driven wind can increase outburst luminosities (see \autoref{wind}) above the predictions of our models, which do not track ejected material.

Five apparently quiescent progenitors of hydrogen-poor SNe have currently been detected in pre-explosion imaging by the {\it Hubble Space Telescope}. These include the type IIb progenitors of SN 2008ax \citep{crockett:08,folatelli:15}, SN 2011dh \citep{vandyk:11,maund:11}, SN 2013df \citep{vandyk:14}, SN 2016gkg \citep{kilpatrick:17,tartaglia:17}, and the type Ib progenitor of iPTF13bvn \citep{cao:13,eldridge:16}. All of these had luminosities $L \! \lesssim \! 2 \! \times \! 10^5 \, L_\odot$ at the time of pre-explosion imaging, and they were likely not outbursting. However, we note that the pre-explosion {\it HST} imaging occurred more than 6 years before each of these SNe, prior to the onset of wave-driven outbursts in our models. Thus, we would not necessarily expect any of those progenitors to have appeared abnormal.

Constraints on progenitor variability for hydrogen-poor SNe are also provided from ground-based data. From PTF data, there is little evidence for luminous outbursts from type IIb progenitors \citep{strotjohann:15}, though in most cases our predicted outburst luminosities are below detection thresholds of $\sim \! 10^7 \, L_\odot$. SN 2012cs may have had an outburst of luminosity $\sim \! 2 \! \times \! 10^6 \, L_\odot$ roughly 680 days before an explosion \citep{strotjohann:15}, which is similar to (though still slightly brighter than) our predicted outbursts. In contrast, hydrogen-rich type IIn supernovae are known to frequently exhibit luminous pre-SN outbursts \citep{gal-yam:09,ofek:13b,ofek:14} similar to outbursts of luminous blue variables. While the timing of these outbursts suggests they could be driven by wave heating, their larger quiescent/outbursting luminosities suggest these outbursts arise from different types of progenitors (likely more massive) than those studied in this paper. Examining wave heating effects in more massive stars, and stars that have undergone a stellar merger \citep{smith:14}, will be needed to determine whether wave heating can contribute to outbursts preceding type IIn supernovae.

The most stringent constraints exist for the progenitor of SN 2011dh \citep{szczygiel:12}, which actually became slightly {\it dimmer} before explosion, but only by a modest factor of $\lesssim \! 20 \%$. The progenitor of type Ic SN 2012fh was also constrained to vary by less than $\sim \! 2500 \, L_\odot$ in optical bands \citep{samson:17}, though we have not yet made predictions for variability of Ic progenitors. In any case, it is evident that luminous $\sim \! 10^6 \, L_\odot$ outbursts like those of our hydrogen-poor models do not occur in all hydrogen-poor SN progenitors. On the other hand, the existence of type Ibn and transitional SNe indicates that greatly enhanced mass loss does occur in a subset of hydrogen-poor SNe. As we shall discuss in \autoref{nonlin}, this may occur because only some progenitor structures allow large amounts of wave energy to escape the core and drive pre-SN outbursts.

\subsection{Wind and Binary Interactions}
\label{wind}

One shortcoming of our method is that it does not track expelled material, instead letting it flow off the edge of our computational grid. Hence, any additional observable effects produced by expelled material will be lost. In particular, we have not accounted for collisions between different shells of material ejected by wave heating during successive outbursts. For instance, in our hydrogen-poor progenitor shown in Figure \ref{fig:Mdotv15Msun_stripped_003}, there are two major phases of mass loss: one during core neon burning, and one during core oxygen burning, each of which eject $\sim 0.05 \, M_\odot$. However, typical ejection speeds are $\sim 50 \, {\rm km}/{\rm s}$ during neon burning, and $\sim 250 \, {\rm km}/{\rm s}$ during oxygen burning. Hence, the shell of material ejected during oxygen burning will catch up with the previously ejected material and collide at a relative velocity of  $\sim 200 \, {\rm km}/{\rm s}$. The collision of these shells will convert $\sim 10^{47} \, {\rm erg}$ of kinetic energy to thermal energy. Spread out over the oxygen burning phase, this will thermalize of order $10^{40} \, {\rm erg}/{\rm s} \approx 3 \times 10^6 \, L_\odot$. A significant fraction of this energy may be radiated in optical bands, similar to interaction powered SNe, and could significantly increase the luminosity of pre-SN outbursts relative to our predictions. Accounting for these shell-shell collisions will be necessary for more reliable predictions of pre-SN outburst luminosities.

Additionally, we have neglected the possible influence of a binary companion, which is likely to be present in the majority of stripped envelope SNe. For orbital periods of hours-weeks, the orbital velocity of the companion will be of order hundreds of ${\rm km}/{\rm s}$, of similar magnitude to our predicted wind velocities. Hence, the interaction between the companion and the wind may be strong, causing some of the wind to accrete onto the companion \citep{mcley:14} or be deflected to form internal shocks. Modeling these effects is beyond the scope of this work, but we speculate that they could significantly increase the radiative luminosity of the wave-driven outbursts.

\section{Wave Non-linearity}
\label{nonlin}

\begin{figure}
\begin{center}
\includegraphics[scale=0.35]{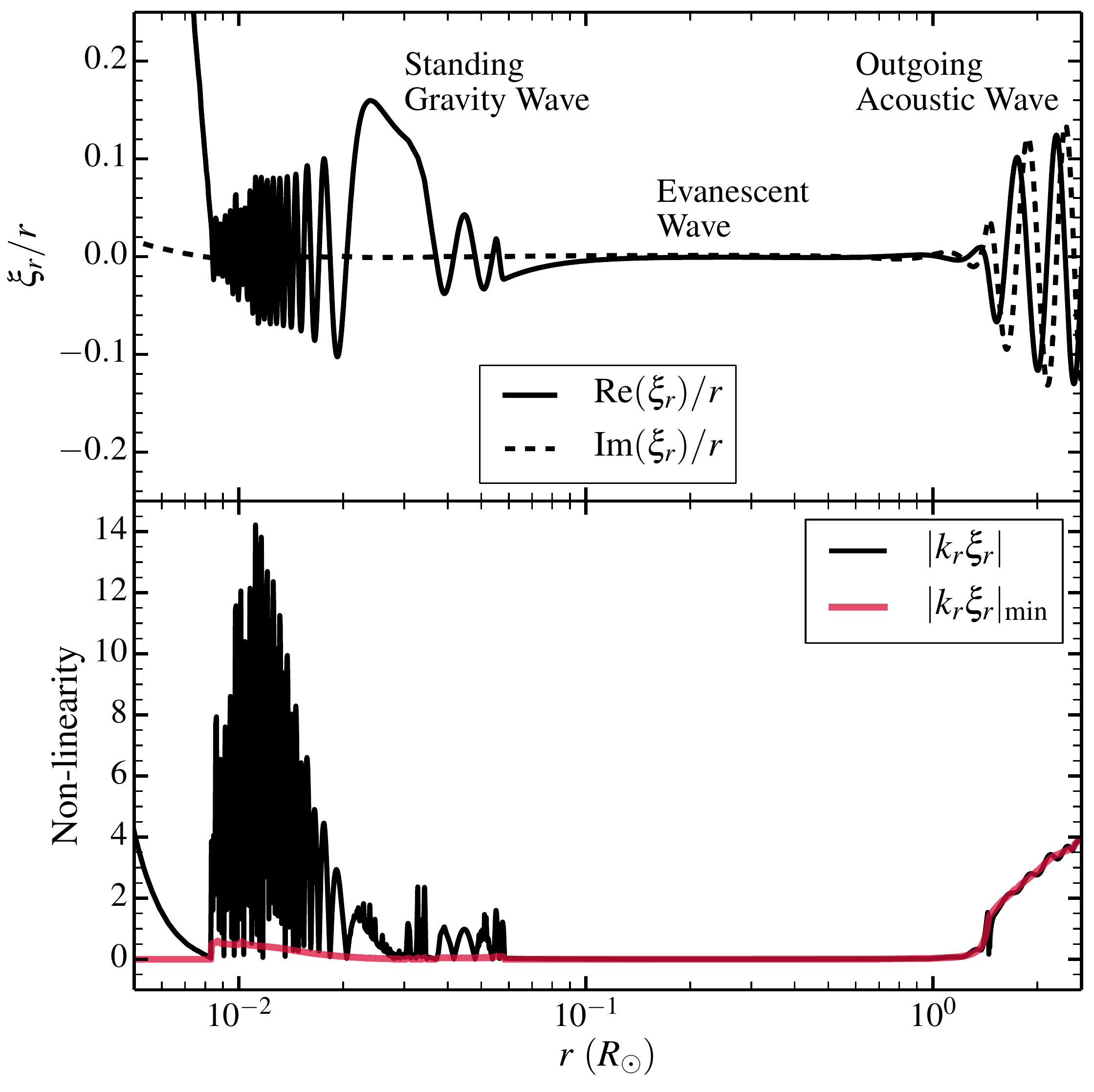}
\end{center} 
\caption{ \label{fig:15MsunWave_stripped} 
{\bf Top:} Real (solid) and imaginary (dashed) components of the radial wave displacement $\xi_r$ during core oxygen burning of our hydrogen-poor model, required to produce an outgoing energy flux of $L_{\rm heat} = 10^7 \, L_\odot$. The imaginary component is small in the core where the waves are standing gravity waves, but it becomes large in the envelope where the waves become outgoing acoustic waves. {\bf Bottom:} Wave non-linearity $k_r \xi_r$. The waves are highly non-linear and subject to dissipative instabilities where $k_r \xi_r \gtrsim 1$. The severe non-linearity in the gravity wave cavity will likely cause most wave energy to be deposited there before escaping to heat the stellar envelope. The red line shows the minimum non-linearity required for waves to carry $L_{\rm heat}$, and represents the wave non-linearity in the absence of wave reflection. }
\end{figure}

Wave-driven outbursts may only occur in some progenitors because of non-linear damping of gravity waves in the radiative outer core. After being  excited at the interface between the convective core and the radiative outer core, gravity waves propagate outward until they encounter an evanescent region, usually created by the convective helium-burning shell (see \autoref{fig:15MsunProp_stripped}). In order to escape the core as acoustic waves that heat the envelope, they must tunnel through this evanescent region, throughout which their amplitude is attenuated by a factor $\sim \! (r_{\rm in}/r_{\rm out})^{\sqrt{l(l+1)}}$, where $l$ is the angular wavenumber and $r_{\rm in}$/$r_{\rm out}$ are the radial coordinates of the inner/outer boundaries of the evanescent region. In our models, the helium burning shell is thick, with $r_{\rm out}/r_{\rm in} \! \sim \! 5$ for typical wave frequencies $\omega_{\rm con}$. Consequently, waves are efficiently reflected at the convective boundary, and only a small fraction of their energy tunnels through. Gravity wave energy must build up in the outer core before the energy leakage rate $L_{\rm heat}$ is equal to the energy input rate by convection. This requires large, potentially non-linear wave amplitudes.

We generally expect waves to become non-linear in the stellar envelope, because the falling density requires wave amplitudes to become large to carry the same energy flux. In the absence of background flows, the wave energy flux is 
\begin{equation}
L_{\rm wave} = 2 \rho r^2 v_g \omega^2 |{\mathbf{\xi}}|^2 \,,
\end{equation}
where $\xi$ is the wave displacement vector and $v_g$ is the radial component of the wave group velocity. Thus, in order to carry a given energy flux, the wave displacement scales as 
\begin{equation}
|\xi| = \bigg[ \frac{L_{\rm wave}}{2 \rho r^2 v_g \omega^2} \bigg]^{1/2} \,.
\end{equation}
A useful measure of the non-linearity of the waves is the product of radial wavenumber and radial wave displacement, $|k_r \xi_r|$. Where $|k_r \xi_r| \! \ll 1 \!$, the waves are nearly linear, while $|k_r \xi_r| \! \geq 1 \!$ implies very non-linear waves. Using the dispersion relations in \autoref{prop}, we find
\begin{equation}
|k_r \xi_r| \simeq  \bigg[ \frac{L_{\rm wave} \, \ell^3 N}{2 \rho r^5 \omega^4} \bigg]^{1/2} \,,
\label{kxiigw}
\end{equation}
while for acoustic waves, 
\begin{equation}
|k_r \xi_r| \simeq \bigg[ \frac{L_{\rm wave}}{2 \rho r^2 c_s^3} \bigg]^{1/2} \,.
\label{kxiac}
\end{equation}
Equations \ref{kxiigw} and \ref{kxiac} are the wave amplitudes given a pure outward going wave. They are lower limits to the required wave amplitude because reflection adds a standing wave component in addition to the traveling wave component. \autoref{fig:15MsunWave_stripped} shows a plot of the minimum wave non-linearity required to carry $L_{\rm wave} = 10^7 \, L_\odot$ within our model, given a wave frequency $\omega= \omega_{\rm con}$. We plot equation \ref{kxiigw} within g mode cavities and equation \ref{kxiac} in evanescent/acoustic cavities. IGW within the core are in principle able to carry $\sim \! 10^7 \, L_\odot$ without becoming strongly non-linear, if they are not reflected and prevented from propagating outward. Note this statement is dependent on the wave frequency $\omega$, and higher frequency IGW can carry more energy. In contrast, the low envelope density causes acoustic waves carrying the same flux to become non-linear as they propagate toward the surface. Hence, in the absence of wave reflection (i.e., no thick intervening evanescent zones), convectively excited IGW can carry large amounts of energy outward, but after tunneling into acoustic waves they become non-linear and steepen into shocks in the stellar envelope.

In real stars, however, evanescent zones cause waves to reflect on their journey outward, bottling up their energy and  increasing the wave amplitude within the core. To calculate realistic wave amplitudes for our models, we solve the linear, forced wave equations within our stellar model shown in Figure \ref{fig:15MsunProp_stripped}. We employ an outgoing acoustic wave boundary condition at $r = 3 \, R_\odot$, and we add a forced inner boundary condition to excite waves. We then normalize the wave function so that the outgoing energy flux is equal to the wave heating rate of $L_{\rm heat} \! \approx 10^7 \, L_\odot$ calculated for that model. \autoref{fig:15MsunWave_stripped} shows the wave function of the radial wave displacement, $\xi_r$, as a function of radius within our model. We see two wave cavities: the gravity wave cavity in between the convective oxygen-burning core and the convective helium-burning shell, and the acoustic wave cavity above the helium-burning shell. The waves are evanescent in most of the intervening helium-burning shell. Because of the thick evanescent region, the waves are mostly reflected at the gravity wave cavity boundaries, producing a standing gravity wave (i.e., a gravity mode) in the outer core. In the acoustic cavity, the waves are nearly pure outgoing waves.

The bottom panel of \autoref{fig:15MsunWave_stripped} shows the corresponding wave non-linearity, $|k_r \xi_r|$. We see that our computed wave function is highly non-linear in both the acoustic and gravity wave cavities. In the acoustic cavity, non-linearity leads to formation of shocks which dissipates wave energy, which is accounted for as discussed in \autoref{shocks}. In the gravity wave cavity, non-linearity likely leads to a turbulent cascade that damps wave energy, which is not accounted for in our models. Hence, in stars containing thick helium-burning shells that hinder wave energy from escaping to the envelope, non-linear dissipation within the outer core may damp the majority of wave energy in the core before it can escape to the envelope. More rigorous studies of non-linear coupling such as \cite{weinberg:08} and \cite{kumar:96} will be needed to determine the wave saturation amplitude in the core and resultant heating of the envelope, but non-linear saturation could cap core g mode amplitudes at least one order of magnitude smaller than our estimates, such that $|k_r \xi_r| \lesssim 1$. This could reduce the envelope heating rate by large factors, two orders of magnitude or more, compared to the predictions of our models that negelect non-linear damping. Hence, we infer that non-linear gravity wave damping will prevent large wave-driven outbursts in many stars, especially those with thick evanescent regions created by thick convective helium-burning shells.

As mentioned in Section \ref{progenitors}, only a subset of progenitors are observed/inferred to produce pre-SN outbursts. Non-linear gravity wave damping may be the reason that many progenitors do not exhibit pre-SN outbursts. However, the models of this paper and \cite{fuller:17} have only investigated relatively ``normal" progenitor models derived from $M_{\rm ZAMS} = 15 \, M_\odot$ primaries. The core structure and evolution can be markedly different for stars of different mass (especially low-mass $\sim \! 10 \, M_\odot$ primaries, \citealt{woosleyheger:15}), or stars that have undergone binary mergers \citep{justham:14}. In future work, we plan to explore wave heating in a more diverse set of progenitor stars, to see which have core structures that allow wave energy to escape into the envelope to drive pre-SN outbursts.

\section{Conclusions}
\label{conclusions}

We have investigated the effects of energy transport by gravity/acoustic waves excited during late nuclear burning phases of hydrogen-poor, stripped-envelope SN progenitors. We have investigated two stellar models composed of a $\sim \! 5 \, M_\odot$ helium core evolved from a $15 \, M_\odot$ progenitor stripped of its hydrogen envelope. At the onset of core carbon burning, our hydrogen-free model is a hot and compact helium star and likely represents a typical type Ib SN progenitor, while our hydrogen-poor model is a yellow supergiant similar to type IIb progenitors. We have used the MESA stellar evolution code to calculate wave heating rates of the stellar envelope and perform 1D hydrodynamic simulations of the progenitors from carbon burning to core-collapse.

Similar to the $15 \, M_\odot$ red supergiant model investigated by \cite{fuller:17}, we find that energy transport during core neon and oxygen burning can have a profound effect on the subsequent evolution and appearance of the star. Convecitvely excited gravity waves tunnel into acoustic waves that carry $\sim \! 10^7 \, L_\odot$ of power into the envelope and damp via weak shock formation fairly close to the surface of the star (though still at large optical depth), depositing enough energy to unbind the overlying material. Wave heating thus drives a dense, super-Eddington wind with mass loss rates of $\sim \! 0.1 \, M_\odot/{\rm yr}$ during core neon and oxygen burning. The progenitor has bolometric luminosities of nearly $10^6 \, L_\odot$ during these wave-driven outbursts, though most of the wave energy is used to unbind material and accelerate the wind.

Our hydrogen-free and hydrogen-poor model behave similarly but exhibit key differences. As a result of being more extended, the hydrogen-poor model suffers more mass loss, losing a total of $0.13 \, M_\odot$ compared to the $0.04 \, M_\odot$ lost from the hydrogen-free model. Wind speeds are $\sim 250 \, {\rm km/s}$ in the hydrogen-poor model compared to $\sim 450 \, {\rm km/s}$ in the hydrogen-free model. As a result of its larger opacity per unit mass, the hydrogen-poor model is cooler at all times, staying in the range $5000 \, {\rm K} \! \lesssim \! T_{\rm eff} \! \lesssim \! 10000 \, {\rm K}$. In contrast, the hydrogen-free model is hot ($T_{\rm eff} \! \sim \! 45000 \, {\rm K}$) and compact during quiescence, but during outburst its photosphere moves out into the optically thick wind, with an effective temperature $T_{\rm eff} \approx 10000 \, {\rm K}$. So, despite similar luminosities in optical bands during outburst, the change in visual band luminosity will be larger for the hydrogen-free model because it radiates primarily in the ultraviolet during quiescence. In both cases, outbursts should be detectable for sufficiently close progenitors like those monitored by the LBT survey \citep{kochanek:08}, but our models do not reach the $\sim \! 10^7 \, L_\odot$ luminosities of pre-SN outbursts like those of SN 2009ip.

However, by explicitly calculating a hydrodynamic wave-form at the estimated wave heating rates, we have shown that gravity waves in the core of the star can become highly non-linear. The non-linearity occurs if the gravity waves are trapped by a thick overlying convective helium-burning shell, such that their energy builds up in the outer core before the waves can tunnel into the envelope as acoustic waves. The likely consequence is that the majority of wave energy will damp via non-linear instabilities in the outer radiative core, where the wave energy is insufficient to drive an outburst or enhanced mass loss. Hence, wave-driven outbursts are probably quenched by non-linear instabilities in many SN progenitors (similar to the non-linear quenching described in \citealt{weinberg:08}), and we suspect this is the reason some SN progenitors, like that of SN 2011dh \cite{szczygiel:12}, did not exhibit pre-SN outbursts.

Nonetheless, pre-SN outbursts do occur in some SN progenitors, and we have shown that the mass loss rates, wind speeds, and outburst durations of our models are approximately what is needed to produce type Ibn SNe (for hydrogen-free progenitors), and transitional IIb/IIn SNe (for hydrogen-poor progenitors). Although our model outburst luminosities are somewhat below those that have been detected, shock heating from wind-wind or wind-binary collisions may increase outburst luminosities. Additionally, the convectively excited wave frequency spectrum and multidimensional effects may affect the results, which we plan to explore in future work.

The question which must now be answered is: which SN progenitors have structures and evolution amenable to wave-driven outbursts, and why? We suspect part of the solution relates to the presence or absence of a convective helium-burning shell: a thick convective shell will insulate the envelope from waves excited in the core, so stars without convective shells overlying the core may be more prone to wave-driven outbursts. Thus far, we have only modeled a few stellar models derived from $15 \, M_\odot$ progenitors. A more extensive survey of the diverse population of SN progenitors will be needed to determine when wave-driven outbursts have their most profound effects on the death throes of massive stars.

\section{Acknowledgments}

JF acknowledges partial support from NSF under grant no. AST-1205732 and through a Lee DuBridge Fellowship at Caltech. This research was supported in part by the National Science Foundation under Grant No. NSF PHY-1125915, and by the Gordon and Betty Moore Foundation through Grant GBMF5076.

\bibliography{MassiveWavesBib}

\appendix

\section{Massive Star Models with MESA}
\label{models}

The models in this paper are constructed in a similar fashion to those in \cite{fuller:17}, with a few important differences. We use MESA \citep{paxton:11,paxton:13,paxton:15} version 9575 for all models in this paper. For evolution through core helium burning, we use the following inlist:

\begin{verbatim}

!-------------------------------------  MAIN      
initial_mass = 15.
initial_z = 0.02
use_Type2_opacities = .true.
Zbase = 0.02  
      
!------------------------------------- OTHER
use_other_energy = .true. 
      
!------------------------------------  MESH            
mesh_delta_coeff = 0.6
varcontrol_target = 3d-4
logQ_min_limit = -20d0
mesh_delta_coeff_for_highT = 1.7
dX_div_X_limit_min_X = 1d-5
dX_div_X_limit = 1d-1
dX_nuc_drop_min_X_limit = 3d-5
dX_nuc_drop_limit = 3d-3
delta_HR_limit = 0.05
delta_HR_hard_limit = 0.1

!------------------------------------  WIND

hot_wind_scheme = 'Dutch'
cool_wind_RGB_scheme = 'Dutch'
cool_wind_AGB_scheme = 'Dutch'
RGB_to_AGB_wind_switch = 1d-4
Dutch_scaling_factor = 0.8

!----------------------------_------  MIXING

overshoot_f_above_nonburn_core = 0.025
overshoot_f0_above_nonburn_core = 0.01
overshoot_f_above_nonburn_shell = 0.025
overshoot_f0_above_nonburn_shell = 0.01
overshoot_f_below_nonburn_shell = 0.025
overshoot_f0_below_nonburn_shell = 0.01

overshoot_f_above_burn_h_core = 0.025
overshoot_f0_above_burn_h_core = 0.01
overshoot_f_above_burn_h_shell = 0.025
overshoot_f0_above_burn_h_shell = 0.01
overshoot_f_below_burn_h_shell = 0.025
overshoot_f0_below_burn_h_shell = 0.01

overshoot_f_above_burn_he_core = 0.025
overshoot_f0_above_burn_he_core = 0.01
overshoot_f_above_burn_he_shell = 0.025
overshoot_f0_above_burn_he_shell = 0.01
overshoot_f_below_burn_he_shell = 0.025
overshoot_f0_below_burn_he_shell = 0.01
     
overshoot_f_above_nonburn_core = 0.025
overshoot_f0_above_nonburn_core = 0.01
overshoot_f_above_nonburn_shell = 0.025
overshoot_f0_above_nonburn_shell = 0.01
overshoot_f_below_nonburn_shell = 0.025
overshoot_f0_below_nonburn_shell = 0.01

!Defining a minimum diffusive mixing 
set_min_D_mix = .true.
min_D_mix = 1d2

/ ! end of controls namelist
\end{verbatim}

After the end of core helium burning, but before carbon burning or helium shell burning, we strip off the hydrogen envelope of the hydrogen-free model by adding the following commands to the inlist.
\begin{verbatim}
remove_H_wind_mdot = 1d-1
remove_H_wind_H_mass_limit = 1d-10
\end{verbatim}
For the hydrogen-poor model, we set 
\begin{verbatim}
remove_H_wind_H_mass_limit = 3d-2
\end{verbatim}
In both cases, we also turn off the Dutch wind mass loss used during prior evolution. This allows us to control the hydrogen mass remaining at core-collapse, and to avoid confusion between prescribed mass loss and wave-driven mass loss. For the hydrogen-poor model, we change 
\begin{verbatim}
mixing_length_alpha = 4
\end{verbatim}
to increase the run speed, as lower mixing lengths lead to density inversions near the surface that cause MESA to run slowly. Increasing the mixing length causes the model to be slightly hotter and more compact than it would be otherwise.

Just before carbon burning, we prepare for our hydrodynamical models including wave-driven mass loss. First, we move the surface boundary to a low optical depth with
\begin{verbatim}
&star_job
relax_initial_tau_factor=.true. 
relax_to_this_tau_factor=1d-2
dlogtau_factor=.1
\end{verbatim}
Next, just before the start of core-carbon burning we set the maximum time step to one year and evolve for 25-50 time steps to make sure the models are very close to hydrostatic equilibrium. We then remove the restriction on time step and turn on hydrodynamics with:
\begin{verbatim}
&star_job
change_initial_v_flag = .true.
change_v_flag = .true.
new_v_flag = .true.
remove_surface_by_optical_depth = 7d-3
repeat_remove_surface_for_each_step = .true.
/ ! end of star_job namelist

&controls

!-------------------------------- HYDRO

mlt_accel_g_theta = 1
max_v_div_cs_for_convection = 1d-1
min_T_for_acceleration_limited_conv_velocity=0d0
max_T_for_acceleration_limited_conv_velocity=1d11
max_conv_vel_div_csound = 1d0
okay_to_remesh = .true.
use_ODE_var_eqn_pairing=.true.
use_dvdt_form_of_momentum_eqn=.true.
use_compression_outer_BC=.true.
log_tau_function_weight=50
log_kap_function_weight=50
R_function3_weight = 50
newton_iterations_limit=9
iter_for_resid_tol2=6
tol_residual_norm1=1d-8
tol_max_residual1=1d-7
tiny_corr_coeff_limit=999999
newton_itermin_until_reduce_min_corr_coeff=999999
\end{verbatim}
Our models experience a very small transient when hydrodynamics are turned on, but it is much smaller than subsequent variability induced by wave heating. We note two important features of the hydrodynamics. First, we limit the acceleration of the MLT convective velocity to the local gravitational acceleration, as discussed in \cite{fuller:17}. Second, we do not allow convection when the velocity is larger than $0.1 c_s$. This prevents the growth of an artificial numerical instability that occurs when the outflow sonic point is located in a convective region. We do not believe it strongly affects the outflow mass loss rate or terminal velocity, since convection is unlikely to affect the wind dynamics as discussed in \cite{quataert:16}. Additionally, trial runs without this restriction behave qualitatively similarly, but are numerically less stable due to artificial density perturbations generated at the sonic point which steepen into shocks as they propagate outward.

In most cases, we try to run our models without the use of artificial viscosity. However, when strong shocks approach the photosphere, MESA is forced to take very small time steps and will sometimes crash. When this occurs, we will sometimes backtrack and restart with the addition of artifical viscosity using
\begin{verbatim}
shock_spread_quadratic = 1d-2 
\end{verbatim}
and we note that the use of \verb| shock_spread_linear | tends to cause more problems than it resolves. This artificial viscosity may affect the dynamics of shocks near the photosphere which typically occur near the beginning and end of wave-driven outbursts, so we caution that our results in these regimes (especially at the end of outbursts when multiple fallback shocks occur) are less reliable.

\end{document}